\documentclass[a4paper, amsfonts, amssymb, amsmath, reprint, showkeys, nofootinbib, twoside]{revtex4-1}
\usepackage[english]{babel}
\usepackage[utf8]{inputenc}
\usepackage[colorinlistoftodos, color=green!40, prependcaption]{todonotes}
\usepackage{mathtools}
\usepackage{amsmath}
\usepackage{amssymb}

\usepackage{amsthm}
\usepackage{mathtools}
\usepackage{physics}
\usepackage{xcolor}
\usepackage{graphicx}
\usepackage[left=23mm,right=13mm,top=35mm,columnsep=15pt]{geometry} 
\usepackage{adjustbox}
\usepackage{placeins}
\usepackage[T1]{fontenc}
\usepackage{lipsum}
\usepackage{csquotes}
\usepackage{rotating}

\newcommand\scalemath[2]{\scalebox{#1}{\mbox{\ensuremath{\displaystyle #2}}}}

\usepackage[pdftex, pdftitle={Article}, pdfauthor={Author}]{hyperref} % For hyperlinks in the PDF
\begin{document}

\title{Creating Qubit States with Degenerate Two-level Systems}

\author{Zhuoran Bao}
    \email[Correspondence email address: ]{zhuoran.bao@mail.utoronto.ca}% Your name
    \affiliation{Dept. of Physics, University of Toronto, Toronto, M5S 1A7, Ontario, Canada}
\author{Daniel F. V. James}
    \email[Correspondence email address: ]{
    dfvj@physics.utoronto.ca}% Your name
    \affiliation{Dept. of Physics, University of Toronto, Toronto, M5S 1A7, Ontario, Canada}

\begin{abstract}
A qubit, or quantum bit, is conventionally defined as ``a physical system for storing information that is capable of existing in either of two quantum states or in a superposition of both'' \cite{OED}. In this paper, we examine the simple question of whether two distinct \emph{levels}, each consisting of multiply degenerate sub-states, could serve as a practical quantum bit. Using two fine structure energy levels containing degeneracy as a model. We find that, in the continuous presence of the field, the atom still undergoes Rabi oscillations, which are suitable for quantum gate construction. In addition, we compute the average fidelity in Hadamard gate performance for a single degenerate atom and postulate the required form of two-atom interaction to construct a controlled Z gate.
\end{abstract}

\maketitle

\section{Introduction}

According to the DiVincenzo criteria \cite{DiVincenzo}, a quantum computer is any data processing device that consists of well-defined qubits, for which qubit gates and readout can be performed. However, if a quantum computer is to realize its potential as a truly disruptive technology, it requires the number of qubits to be in the scale of millions \cite{Gidney, Lee, Beverland}; while the current technology delivers qubits on the scale of tens to hundreds \cite{Bohnet, Hawaldar, Kielpinski, Arute, Yulin}. This disparity between the requirements of an idealized quantum computer and current experimental reality has prompted multiple architectural proposals, including but not limited to trapped ions, ultra-cold atom arrays, superconducting qubits, and diamond centers; indeed, this multiplicity by itself implies, with no small emphasis, that no one technology has attained preeminence. Hence, it would always be interesting to examine all possible degrees of freedom available in the system and their capacity for qubit-like behaviours.

In examining various types of quantum computers, the question naturally arises: What are the criteria for an individual quantum system to be considered a qubit? In particular, do they truly have to be a two-state system? When trapped ions are used, some means of removing degeneracies to leave a two-state system is generally employed. In a hyperfine qubit\cite{Harty, Olmschenk, Feng}, the nuclear spin is inherently available for lifting the degeneracy. For Zeeman qubits\cite{Ruster, Brown}, an external magnetic field is necessary to produce a two-state system. However, in the case of an optical qubit\cite{Clark, Mundt}, a magnetic field is applied, but it doesn't suggest a necessity as in the previous two cases. Meanwhile, hyperfine qubits in general involve rare isotopes, and maintaining a stable external magnetic field can pose additional experimental challenges since any fluctuation in such a field induces dephasing. Allowing the degeneracy avoids the problems mentioned above. Additionally, the initialization of the state can be simpler since we do not require all the population to be in a single Zeeman sub-state of the \(^2S_{1/2}\) orbital. Nevertheless, if one chooses to initialize the state in one sub-state of the ground S orbital using the multi-level encoding strategy\cite{Matthew}, then the sub-state populations might also be used for encoding, and a new type of interaction gate, the degenerate Hadamard gate, is now available, providing flexibility for quantum computing codes.

In this manuscript, we consider whether degeneracy lifting is mandatory for optical qubits. As an elementary step, we abandon the traditional notion of a qubit as a two-state system. Instead, we investigate the oscillatory behaviour of the degenerate atomic electron levels and the atom's ability to perform quantum gate operations. We concentrated on the specific example of a two-level system formed by \(^2S_{1/2}\) and \(^2P_{1/2}\) orbitals connected by dipole interaction for simplicity. These two levels are more accessible experimentally for demonstrating Rabi oscillation. A more practical optical qubit could be constructed using the \(^2S_{1/2}\) and \(^2D_{1/2}\) levels, given their long coherence times. In that case, one can either use twisted light to mediate the state transition via a quadrupole interaction or a stimulated Raman transition with a particular polarization and geometry.

The paper is arranged as follows. Section II introduced the notation for Rabi oscillations with degenerate atomic levels connected by dipole interaction. We constructed a degenerate Hadamard gate under the assumption that no external static magnetic field is present. An example for \(^2S_{1/2}\) and \(^2P_{1/2}\) degenerate fine structure states is presented in Appendix A. In section III, we added a weak static magnetic field interaction term to our Hamiltonian. We constructed the time evolution operator under the total Hamiltonian as a power series of the magnetic field. We also computed the average fidelity of the new degenerate Hadamard gate. In section IV, we discussed the requirements for implementing a Controlled Z gate on two degenerate atoms. Section V lists the main findings as the paper's conclusion.

\section{Rabi Oscillation and Single Qubit Gate}
The dipole interaction induces a transition in the atomic state. The Hamiltonian takes the form \cite{AQJ}:
\begin{equation}\label{atomic transition}
     {\hat{\mathcal{H}}} =  {\hat{\mathcal{H}}_0} +  { \hat{\boldsymbol{d}}} \cdot \boldsymbol{E}\cos(\omega t),
\end{equation}
where \( \hat{\mathcal{H}}_0\) describes the two degenerate levels of the atom, \( \hat{ \boldsymbol{d}}\) is the dipole moment, and \( \boldsymbol{E}\cos(\omega t)\) describes the optical field.

Consider a system consisting of two multiply-degenerate levels with states labelled by \(\{\vert g,m_i\rangle\}\) and \(\{\vert e, m_j\rangle\}\), where g and e represent the ground and the excited fine structure energy level. The magnetic quantum numbers \(m_i\) and \(m_j\) are the degeneracy. Any state in the interaction picture can be written as \(\vert \Tilde{\psi}(t)\rangle = \sum_{i} \alpha_i(t)\vert g,m_i\rangle + \sum_{j}\beta_j(t) \vert e,m_j\rangle\). And so, the system's state at any instant, t, is described by a set of differential equations:
\begin{equation}
\label{sys1}
\begin{split}
    i\hslash \frac{\partial}{\partial t}\alpha_i(t) & = \sum_j \beta_j(t) e^{-i\omega_0 t}\Omega_{ij}\frac{1}{2}(e^{i\omega t}+e^{-i\omega t}) \\
    i\hslash \frac{\partial}{\partial t}\beta_j(t) & = \sum_i \alpha_i(t) e^{i\omega_0 t}\Omega_{ji}\frac{1}{2}(e^{i\omega t}+e^{-i\omega t}).
\end{split}
\end{equation}
Here, we use \(\omega_0\) to represent the transition frequency between energy levels g and e. We use \(\Omega_{ij}=\langle g, m_i \vert  { \boldsymbol{d}}\cdot \boldsymbol{E}\vert e,m_j\rangle\) to represent the coupling strength. Using the rotation wave approximation, we would find the solutions for \(\alpha_i(t)\) and \(\beta_j(t)\) correspond to a linear combination of sinusoidal functions of different frequencies. Therefore, we expect the possibility for the state to be at energy level g or e to exhibit a quasiperiodic time dependence (see Figure 1).
\begin{figure}
    \centering
    \includegraphics[width=1\linewidth]{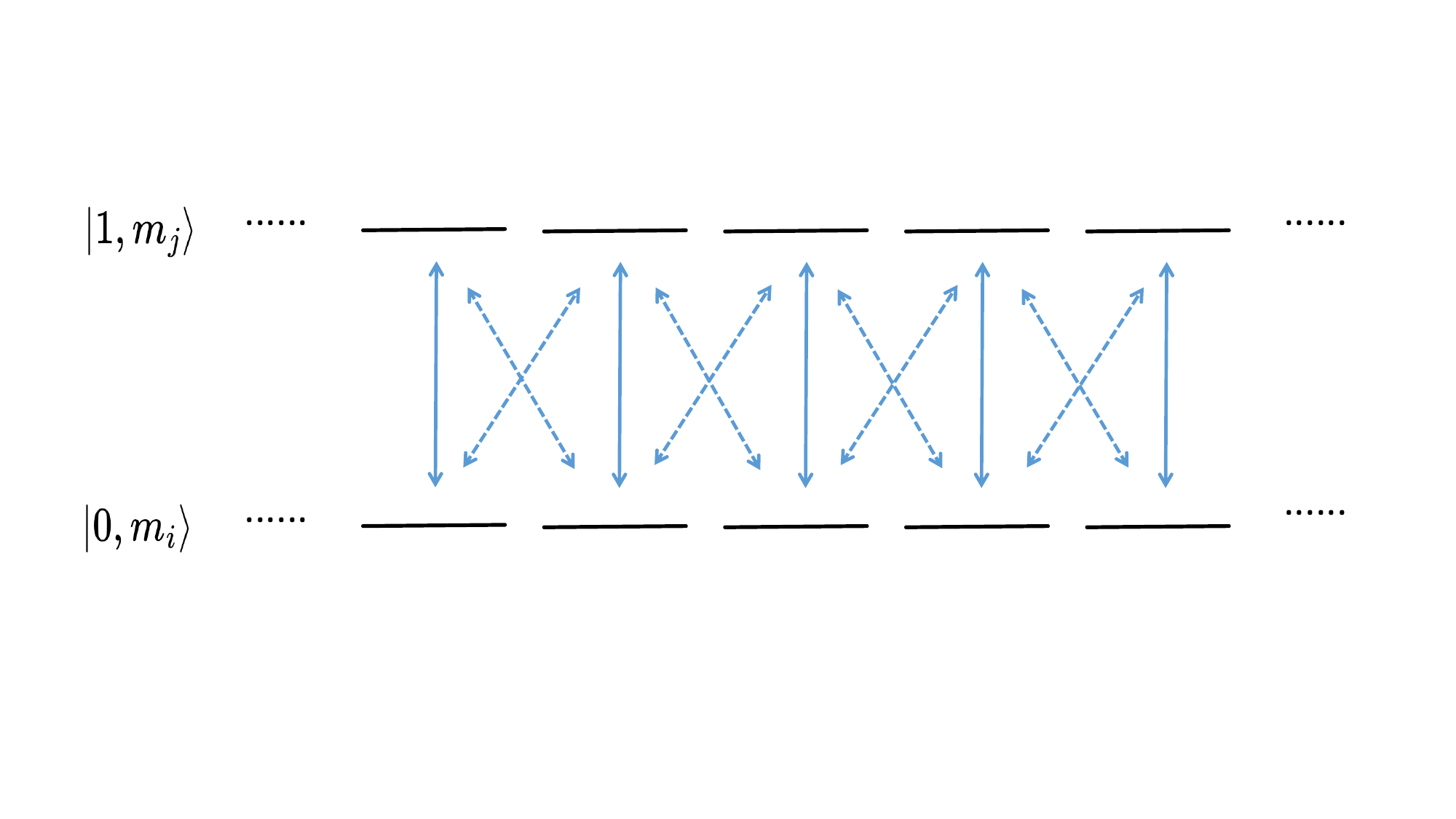}
    \caption{All Possible transitions between the degenerate states in energy levels 0 and 1. Blue solid lines represent the transition mediated by \(\pi_0\), linearly polarized photons, while the dashed lines represent the transition mediated by \(\sigma_{\pm}\), circularly polarized photons.}
    \label{fig:enter-label}
\end{figure}

However, since the atoms are spherically symmetric, additional constraints can be applied to the expectation value of the dipole moment, \(\Omega_{ij}\), based on the angular momentum selection rules. Using 3j-Symbols and the Wigner–Eckart theorem \cite{cowan}:
\begin{widetext}
\begin{equation}\label{o3}
    \Omega_{ij} = \sum_{k=1}^3e \boldsymbol{\epsilon}_k \cdot  \boldsymbol{E} \langle g,m_i \vert  {r_k}\vert e, m_j\rangle 
    = e \langle g \vert \vert rC^{(1)}\vert\vert e \rangle \sum_{k=1}^3 \sum_{q=-1}^1 \begin{pmatrix}
J^{(g)} & 1 & J^{(e)}\\
-m_i & q & m_j
\end{pmatrix} c_k^{(q)}  \boldsymbol{\epsilon}_k \cdot  \boldsymbol{E},
\end{equation}
\end{widetext}
where \(\langle g\vert \vert rC^{(1)}\vert\vert e\rangle\) is a constant that can be related to the Einstein A coefficient for total spontaneous decay \cite{Daniel, cowan}:
\begin{equation}
    \langle g\vert \vert rC^{(1)}\vert\vert e\rangle = \sqrt{\frac{3A_{ge}n}{4c\alpha k_{ge}^2}}.
\end{equation}
The coefficient \(\alpha = e^2/4\pi\epsilon_0c\hbar\) is the fine structure constant, \(k_{ge}\) is the wave number corresponding to the transition between the two energy levels, and n is the number of degeneracy states within the excited energy level. Define \(S=e\langle g\vert \vert rC^{(1)}\vert\vert e\rangle\) which \(S^2\) is sometimes refer to as the atomic line strength \cite{Woodgate}.

Let's consider transitions induced by linearly polarized light. Atoms or ions have spherical symmetry; they have no preferred quantization axis. Therefore, we can define \emph{the polarization direction of the electric field as the z-axis of the atom}. Thereby, we simplify Eq. (\ref{o3}) to:
\begin{equation}
    \Omega_{ij} 
    = S  \begin{pmatrix}
J^{(g)} & 1 & J^{(e)}\\
-m_i & 0 & m_j
\end{pmatrix}  \vert \boldsymbol{E}\vert.
\end{equation}
Angular momentum selection rule \(-m_i+q+m_j=0\) must be satisfied for non-zero coupling constant; Thus, the only allowed transitions are between the pairs, \(\vert g,m_i\rangle\) to \(\vert e, m_i\rangle\) with identical coupling strengths:
\begin{equation}
\label{omega value}
    \Omega_{ii} 
    = S \begin{pmatrix}
J^{(g)} & 1 & J^{(e)}\\
-m_i & 0 & m_i
\end{pmatrix} \vert \boldsymbol{E}\vert.
\end{equation}
The system of equations is decoupled into pairs of coupled equations according to their angular momentum index:
\begin{equation}\label{sys3}
    \begin{split}
    i \frac{\partial}{\partial t}\alpha_i(t) & = \beta_i(t) e^{i(\delta t)}\frac{ \Omega_{ii}}{2\hslash} \\
    i \frac{\partial}{\partial t}\beta_i(t) & = \alpha_i(t) e^{-i(\delta t)}\frac{ \Omega_{ii}^*}{2\hslash}.
    \end{split}
\end{equation}
Each pair can be viewed as a two-state system, hence a qubit. The total state is a superposition of all these two-state systems.

By choosing appropriate energy levels, we can make all Rabi frequencies \(\Omega_{ii}\) in Eq. (\ref{sys3}) equal and real. Therefore, we expect the solution to exhibit Rabi oscillations, and there is no coupling between pairs of degenerate states with different indices \(m_i\) (see Figure 2). In other words, the Hilbert space is decomposed into superpositions of identical rank-two subspaces, and the transition operations are identically applied to each subspace.
\begin{figure}
    \centering
    \includegraphics[width=1\linewidth]{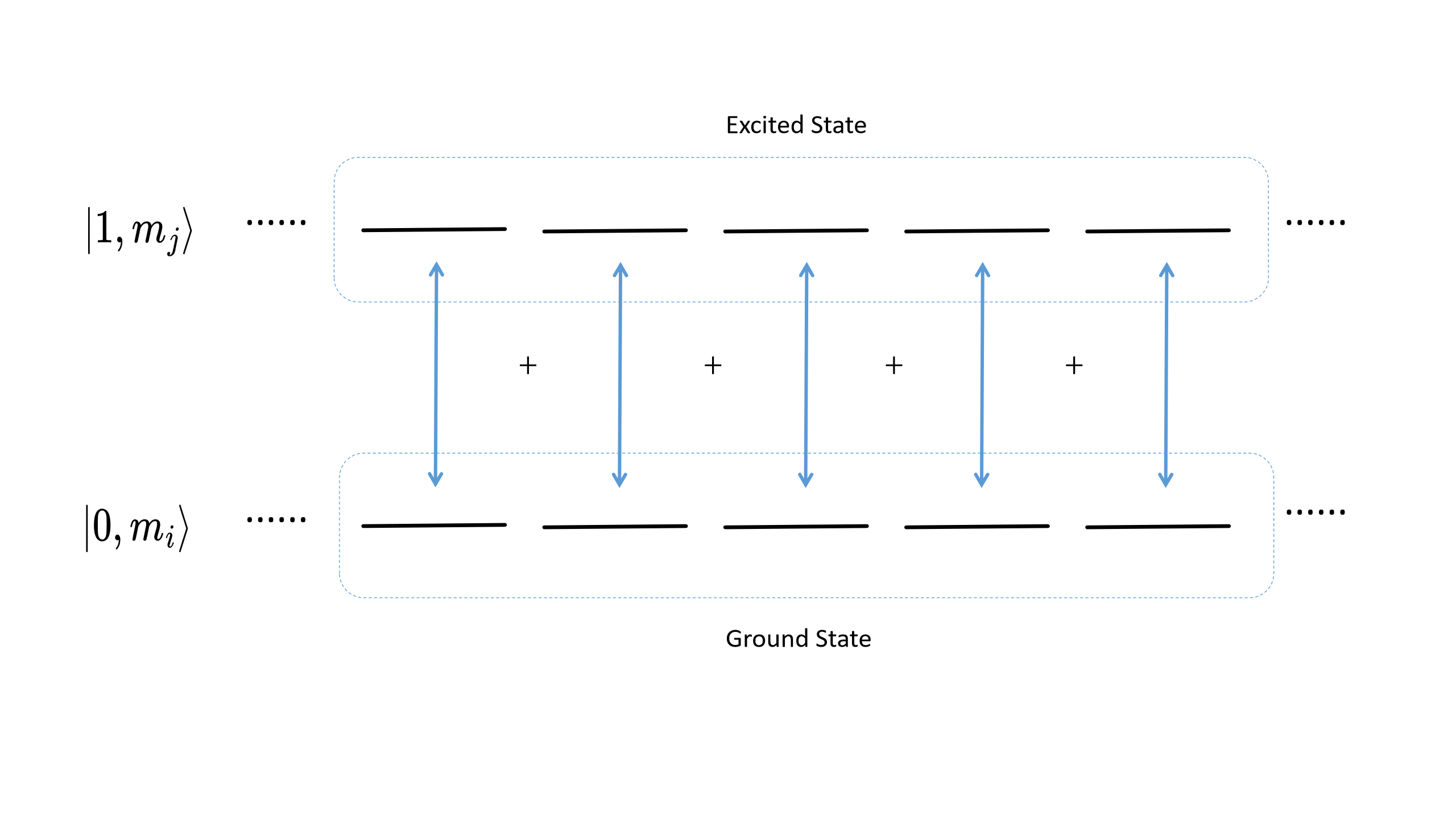}
    \includegraphics[width=1\linewidth]{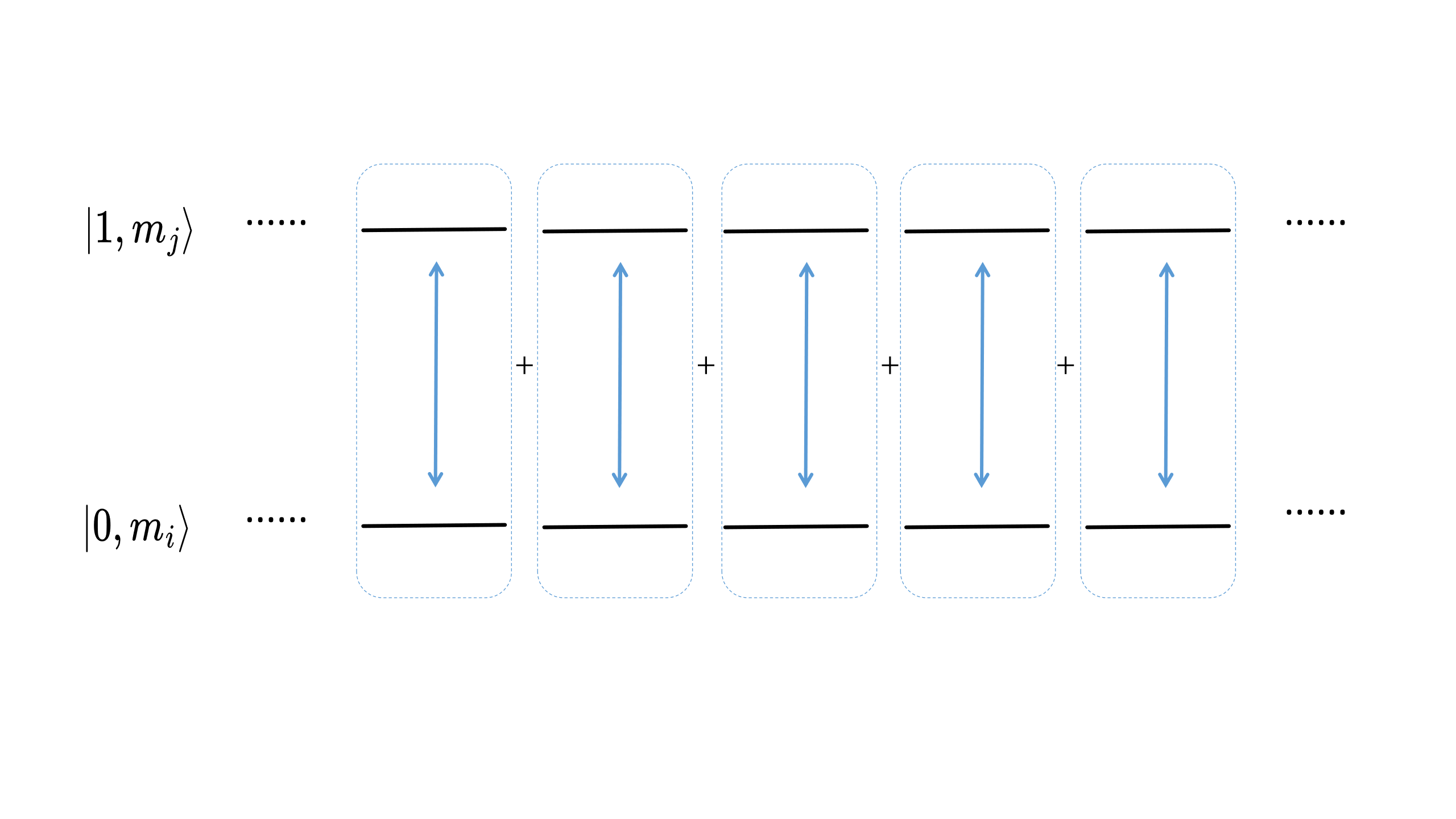}
    \caption{By using linearly polarized light and defining the polarization direction of the electromagnetic field as the z-direction of the atom, only one Rabi frequency remains.}
    \label{fig:enter-label}
\end{figure}
A concrete example for constructing a degenerate Hadamard gate between \(^2S_{1/2}\) and \(^2P_{1/2}\) fine structure energy levels is presented in Appendix A. The degenerate Hadamard gate is given by Had:
\begin{equation}
Had = \frac{1}{\sqrt{2}}\begin{pmatrix}
    1&0&1&0\\
    0&1&0&1\\
    1&0&-1&0\\
    0&1&0&-1
\end{pmatrix},
\end{equation}
where the entries in the first and the third row correspond to the couplings between the \(\vert \frac{1}{2},-\frac{1}{2}\rangle\) degenerate states in the \(^2S_{1/2}\) and \(^2P_{1/2}\) orbital. The entries in the second and fourth rows correspond to the couplings between \(\vert \frac{1}{2},\frac{1}{2}\rangle\) degenerate states in the \(^2S_{1/2}\) and \(^2P_{1/2}\) orbital. We may conclude that generic systems with spherical symmetry can exhibit qubit-like behaviour, provided the interaction term breaks symmetry only in one direction.

\section{The Fidelity of the Degenerate Hadamard Gate with the presence of a weak static Magnetic Field}
A static magnetic field breaks the spherical symmetry of an atom, resulting in the Zeeman effect. The same electric field discussed previously may couple degenerate states with different magnetic quantum numbers \(m_i\) depending on its polarization. The perfect condition described in part one is likely impractical. Therefore, we would like to compute the fidelity for performing the degenerate Hadamard gate in the presence of the static magnetic field.

The usual description of the Zeeman effect proceeds as follows. First, the quantization axis is chosen to be the direction of the static magnetic field. Conventionally, we denote the quantization axis to be the z-axis. Then, the polarization of the electric field that induces the transition is assumed to be arbitrary and decomposed into components along the axis. Finally, the matrix elements for the electric dipole transition are calculated. However, this description does not suit our analysis, as it cannot provide a series expansion for the corrections in the degenerate Hadamard gate as a function of the magnetic field strength. To remedy this, we derived the equivalence of the Zeeman effect expression, which treats the electric polarization as the z-axis while letting the magnetic field direction be arbitrary.

The effective Hamiltonian for the atomic electrons interacting with an external magnetic field \(\boldsymbol{B}\) is:
\begin{equation}\label{H_B1}
    \hat{\mathcal{H}}_B = \mu_B(\hat{\boldsymbol{L}}+g_s\hat{\boldsymbol{S}})\cdot\boldsymbol{B}.
\end{equation}
Where \(\mu_B\) is the Bohr magneton, \(\hat{\boldsymbol{L}}\) is the orbital angular momentum operator, \(\hat{\boldsymbol{S}}\) is the spin angular momentum operator, and \(\boldsymbol{B}\) is the magnetic field vector. Define the direction of electric field polarization as the z-direction. We can write \(\boldsymbol{B}=\boldsymbol{B}_{\parallel z}+\boldsymbol{B}_{\perp z}\). Let the angle between the electric field polarization and the magnetic field be \(\theta\). Without loss of generality, we can define the direction of \(\boldsymbol{B}_{\perp z}\) to be the x-axis. Hence, Eq. (\ref{H_B1}) can be written as:
\begin{widetext}
\begin{equation}
    \hat{\mathcal{H}}_B = \mu_B(\hat{L}_z+g_s\hat{S}_z)\cos{(\theta)}\vert\boldsymbol{B}\vert+\mu_B(\hat{L}_x+g_s\hat{S}_x)\sin{(\theta)}\vert\boldsymbol{B}\vert.
\end{equation}
Using the fine structure level labelled by \(\gamma L SJ\) with degeneracy \(m_J\) as a basis, \(\hat{\mathcal{H}}_B\) can be represented as a matrix. With the principle quantum number \(\gamma\) unchanged, each matrix element of \(\hat{\mathcal{H}}_B\) is given by \(\langle \gamma L'S'J'm_J'\vert \hat{\mathcal{H}}_B\vert \gamma LSJm_J\rangle\). To evaluate the matrix element, use the projection theorem and follow the derivation in \cite{cowan, Woodgate}:
\begin{equation}\label{projection L}
    \langle\gamma LSJm_j\vert \hat{L}_i\vert\gamma L'S'J'm_j'\rangle = \frac{\langle\gamma LSJm_j\vert \hat{\boldsymbol{L}}\cdot\hat{\boldsymbol{J}}\vert\gamma LSJm_j\rangle}{\hbar j(j+1)}\langle\gamma LSJm_j\vert \hat{J}_i\vert\gamma L'S'J'm_j'\rangle,
\end{equation}
\begin{equation}
    \langle\gamma LSJm_j\vert \hat{S}_i\vert\gamma L'S'J'm_j'\rangle = \frac{\langle\gamma LSJm_j\vert \hat{\boldsymbol{S}}\cdot\hat{\boldsymbol{J}}\vert\gamma LSJm_j\rangle}{\hbar j(j+1)}\langle\gamma LSJm_j\vert \hat{J}_i\vert\gamma L'S'J'm_j'\rangle,
\end{equation}
\end{widetext}
By diagonalizing \(\hat{\mathcal{H}}_0+\hat{\mathcal{H}}_{dipole}+\hat{\mathcal{H}}_B\) in the \(\vert \gamma LSJm_j\rangle\) basis, we can find the time evolution operator and its series expansion with respect to magnetic field strength. Consider the \(^2S_{1/2}\) to \(^2P_{1/2}\) transition, \(\hat{\mathcal{H}}_B\) as a matrix is:
\begin{equation}
\begin{split}
    \hat{\mathcal{H}}_B & = \frac{\mu_B g_s\vert\boldsymbol{B}\vert}{2}\left(
\begin{array}{cccc}
 -\cos (\theta) & \sin (\theta) & 0 & 0 \\
 \sin (\theta) & \cos (\theta) & 0 & 0 \\
 0 & 0 & -\frac{\cos (\theta)}{3} & \frac{\sin (\theta)}{3} \\
 0 & 0 & \frac{\sin (\theta)}{3} & \frac{\cos (\theta)}{3} \\
\end{array}
\right) \\
\end{split}
\end{equation}
Let \(g_s\vert\boldsymbol{B}\vert = B_0\) for simplicity. Note that \(\hat{\mathcal{H}}_B\) remains stationary in the interaction picture; the total Hamiltonian in the interaction picture is:
\begin{widetext}
\begin{equation}
\hat{\mathcal{H}}=\hat{\mathcal{H}}_{int}+\hat{\mathcal{H}}_B =\frac{\hbar\Omega}{2}\begin{pmatrix}
        0&0&1&0\\
        0&0&0&1\\
        1&0&0&0\\
        0&1&0&0
    \end{pmatrix} +\frac{\mu_BB_0}{2}\left(
\begin{array}{cccc}
 -\cos (\theta) & \sin (\theta) & 0 & 0 \\
 \sin (\theta) & \cos (\theta) & 0 & 0 \\
 0 & 0 & -\frac{\cos (\theta)}{3} & \frac{\sin (\theta)}{3} \\
 0 & 0 & \frac{\sin (\theta)}{3} & \frac{\cos (\theta)}{3} \\
\end{array}
\right).
\end{equation}
\end{widetext}
To treat the magnetic field as a perturbation compared to the dipole transition, we require that \(\mu_BB_0/\hbar\Omega \ll 1\). 

We find the eigenvalues and eigenvectors of the total Hamiltonian and construct the time evolution operator \(\hat{U}^{(int)}_{tot}(t)\). In the Schrodinger picture, \(\hat{U}_{tot}(t)=\hat{U}_0(t)\hat{U}_{tot}^{(int)}(t)\) where \(\hat{U}_0(t)\) is the time evolution given by \(\hat{\mathcal{H}}_0\) in Eq. (\ref{atomic transition}) representing the Hamiltonian of the atomic electron without any external fields. To construct the degenerate Hadamard gate, the dipole-free time evolution of the state is no longer \(\hat{U}_0(t)\) because of the presence of the magnetic field. Instead, it is the time evolution of the Hamiltonian \(\hat{\mathcal{H}}_0+\hat{\mathcal{H}}_B\). Define this time evolution operator as \(\hat{U}_0' (t)\). By finding the eigenvalue and eigenvector of \(\hat{\mathcal{H}}_0+\hat{\mathcal{H}}_B\), we are able to compute \(\hat{U}'(t)\). The following sequence of operations implements the degenerate Hadamard gate: \(\hat{U}_{Had}=\hat{U}' (7\pi/2\omega-\pi/2\Omega)U(\pi /2\Omega)U' (3 \pi/2\omega)\) with an extra global phase of i. We expanded the degenerate Hadamard gate as a power series of \(\mu_BB_0/\hbar\Omega\). The zeroth order recovers the degenerate Hadamard gate, and higher orders of the expansion are proportional to powers of \(\mu_BB_0/(\hbar\Omega) \ll 1\) as expected (Higher ordered terms see appendix E):
\begin{equation}\label{U0Had}
    \hat{U}_{Had}^{(0)} =\frac{1}{\sqrt{2}}\left(
\begin{array}{cccc}
 1 & 0 & 1 & 0 \\
 0 & 1 & 0 & 1 \\
 1 & 0 & -1 & 0 \\
 0 & 1 & 0 & -1 \\
\end{array}
\right).
\end{equation}

Next, we would like to calculate the average fidelity of the degenerate Hadamard gate. Let \(\hat{H}\) denote the degenerate Hadamard gate as in Eq. (\ref{U0Had}). Consider only the case of comparing two unitary operators acting on pure states, and suppose the Hilbert space is n-dimensional, we can define the average fidelity \cite{Pedersen}:
\begin{equation}
    \int_{S^{2n-1}}\langle\psi\vert (\hat{H}) (\hat{U}_{Had})\vert \psi\rangle d\sigma_{\psi}/V_{S^{2n-1}},
\end{equation}
where the states are represented as points on the surface of a unit sphere in 2n dimensions, \(d\sigma_{\psi}\) is the area element, and \(V_{S^{2n-1}}\) is the surface area of the 2n-1 unit sphere. It has been shown that \cite{Pedersen}:
\begin{widetext}
\begin{equation}\label{fidelity}
\begin{split}
    \int_{S^{2n-1}} \vert\langle\psi\vert (\hat{H}) (\hat{U}_{Had})\vert \psi\rangle \vert ^2 d\sigma_{\psi}/V_{S^{2n-1}} &= \frac{1}{n(n+1)}[Tr((\hat{H})(\hat{U}_{Had})(\hat{U}_{Had})^{\dag}(\hat{H})^{\dag})+\vert Tr((\hat{H})(\hat{U}_{Had})\vert^2] \\
    & = \frac{1}{n(n+1)}[n+\vert Tr((\hat{H})(\hat{U}_{Had}))\vert^2].
\end{split}
\end{equation}
\end{widetext}
The above integral can be computed using Corollary 3.13 in \cite{Zhang}. A case-specific proof is also presented in Appendix D. Using the power series of \(\hat{U}_{Had}\) from before and Eq. (\ref{fidelity}), we find that the average fidelity of the degenerate Hadamard gate is also a power series of \(\mu_BB_0/(\hbar\Omega)\). Let \(\omega\) be the transition frequency without the presence of the magnetic field. Keep to the 2nd order, the fidelity of the degenerate Hadamard gate is:
\begin{widetext}
\begin{equation}
    F(\hat{U}_{Had})_{avg} =1-\frac{(\mu_BB_0)^2}{(\hbar\Omega)^2}\frac{458 \pi ^2 (\Omega/\omega)^2+2 \pi (20-7 \pi )  (\Omega/\omega)+\left(8-4 \pi +\pi ^2\right)}{180}\ . 
\end{equation}
\end{widetext}
The average fidelity of the degenerate Hadamard gate depends on the value of \(\mu_BB_0/(\hbar\Omega)\) as expected. For the error to be negligible, we require the magnetic field \(B_0\ll \hbar\Omega/\mu_B\). The exact bound on the magnetic field depends on the transition's Rabi frequency. To roughly estimate the bound, we consider a Rabi oscillation frequency in the range of \(10^3\)Hz to \(10^6\)Hz, then we require the magnetic field to be much smaller than \(10^{-8}\)T to \(10^{-5}\)T. Shielding of external field using \(\mu\) metal should be feasible in a lab \cite{Ruster}.

\section{Controlled Z Gate between two Qubits}
The power of quantum computation lies in the ability to perform two-qubit gates. Here, we generalized the approach given in \cite{AQJ} to degenerate systems. Using the same strategy presented in the reference, we attempt to set limitations on the interaction Hamiltonian such that the time evolution of the combined system takes the form:
\begin{widetext}
\begin{equation}
\label{two atom unitary}
    \hat{U}(t) = \left(
\begin{array}{cccccccccccccccc}
 e^{i a t} & 0 & 0 & 0 & 0 & 0 & 0 & 0 & 0 & 0 & 0 & 0 & 0 & 0 & 0 & 0 \\
 0 & e^{i a t} & 0 & 0 & 0 & 0 & 0 & 0 & 0 & 0 & 0 & 0 & 0 & 0 & 0 & 0 \\
 0 & 0 & c^* & 0 & 0 & 0 & 0 & 0 & d & 0 & 0 & 0 & 0 & 0 & 0 & 0 \\
 0 & 0 & 0 & c^* & 0 & 0 & 0 & 0 & 0 & d & 0 & 0 & 0 & 0 & 0 & 0 \\
 0 & 0 & 0 & 0 & e^{i a t} & 0 & 0 & 0 & 0 & 0 & 0 & 0 & 0 & 0 & 0 & 0 \\
 0 & 0 & 0 & 0 & 0 & e^{i a t} & 0 & 0 & 0 & 0 & 0 & 0 & 0 & 0 & 0 & 0 \\
 0 & 0 & 0 & 0 & 0 & 0 & c^* & 0 & 0 & 0 & 0 & 0 & d & 0 & 0 & 0 \\
 0 & 0 & 0 & 0 & 0 & 0 & 0 & c^* & 0 & 0 & 0 & 0 & 0 & d & 0 & 0 \\
 0 & 0 & -d^* & 0 & 0 & 0 & 0 & 0 & c & 0 & 0 & 0 & 0 & 0 & 0 & 0 \\
 0 & 0 & 0 & -d^* & 0 & 0 & 0 & 0 & 0 & c & 0 & 0 & 0 & 0 & 0 & 0 \\
 0 & 0 & 0 & 0 & 0 & 0 & 0 & 0 & 0 & 0 & e^{i a t} & 0 & 0 & 0 & 0 & 0 \\
 0 & 0 & 0 & 0 & 0 & 0 & 0 & 0 & 0 & 0 & 0 & e^{i a t} & 0 & 0 & 0 & 0 \\
 0 & 0 & 0 & 0 & 0 & 0 & -d^* & 0 & 0 & 0 & 0 & 0 & c & 0 & 0 & 0 \\
 0 & 0 & 0 & 0 & 0 & 0 & 0 & -d^* & 0 & 0 & 0 & 0 & 0 & c & 0 & 0 \\
 0 & 0 & 0 & 0 & 0 & 0 & 0 & 0 & 0 & 0 & 0 & 0 & 0 & 0 & e^{i a t} & 0 \\
 0 & 0 & 0 & 0 & 0 & 0 & 0 & 0 & 0 & 0 & 0 & 0 & 0 & 0 & 0 & e^{i a t} \\
\end{array}
\right),
\end{equation}
\end{widetext}
where c and h are arbitrary complex functions of time. We examine the assumptions to produce the above time evolution and show that this unitary evolution is sufficient to construct a CZ gate.

We start by considering the interaction between two atoms, each degenerate in the \(^2S_{1/2}\) ground and \(^2P_{1/2}\) excited levels. We label the basis for the first atom to be \(\vert \alpha_{i}\rangle\), and for the second atom, \(\vert \beta_{j}\rangle\), where \(i,j=0,1,2,3\) correspond to the ground and the excited energy level with degeneracy. Then, the combined basis is \(\vert \alpha_{i}\beta_{j}\rangle\). Let \(\hat{\mathcal{H}}_{AB}\) be the Hamiltonian describe the two atoms system, then \(\hat{\mathcal{H}}_{AB}\) can be represented in the given basis \(\vert \alpha_{i}\beta_{j}\rangle\) with matrix elements given by:
\begin{equation}\label{hmn}
    h_{mn}= \langle \alpha_i \beta_j \vert \hat{\mathcal{H}} \vert \alpha_k \beta_l\rangle,
\end{equation}
with \(m = 4i+j, n=4k+l\). 
The matrix \(\hat{\mathcal{H}}_{AB}\) can be decomposed as:
\begin{equation}
    \hat{\mathcal{H}}_{AB} = \hat{\mathcal{H}}_{A}\otimes \hat{I} + \hat{I} \otimes \hat{\mathcal{H}}_{B}+\hat{V}_{AB},
\end{equation}
where \(\hat{I}\) is the identity matrix, \(\hat{\mathcal{H}}_{A}\) and \(\hat{\mathcal{H}}_{B}\) are matrices representing the individual Hamiltonian for the first and the second atom in the basis \(\vert\alpha_i\rangle, \vert\beta_j\rangle\) correspondingly, and \(\hat{V}_{AB}\) is the matrix represent the interaction between the two atoms.
In our chosen basis, assuming the two atoms are identical, \(\hat{\mathcal{H}}_A\) and \(\hat{\mathcal{H}}_B\) take the form:
\begin{equation}
    \hat{\mathcal{H}}_A = \hat{\mathcal{H}}_B = \frac{\hbar\omega}{2}\begin{pmatrix}
        1 & 0 &0&0\\
        0& 1 &0&0\\
        0&0&-1&0\\
        0&0&0&-1
    \end{pmatrix},
\end{equation}
with \(\omega = (E_g-E_e)/\hbar\). The interaction \(\hat{V}_{AB}\) is given by:
\begin{equation}
    \hat{V}_{AB} =\hat{\mathcal{H}}_{AB}-\hat{\mathcal{H}}_{A}\otimes \hat{I} - \hat{I} \otimes \hat{\mathcal{H}}_{B}.
\end{equation}
Explicitly, \((\hat{V}_{AB})_{ii} = h_{ii}-\hbar \omega/2\) for i = \{0,1,...,7\}, \((\hat{V}_{AB})_{ii} = h_{ii}+\hbar \omega/2\) for i = \{8,9,...,15\}, and \((\hat{V}_{AB})_{ij} = h_{ij}\) for \(i \neq j\). Now, we would like to consider the meaning of each interaction term and find the required assumption that Eq. (\ref{two atom unitary}) is true. 

Since we assume the two atoms are identical, the diagonal entries of the total Hamiltonian, Eq. (\ref{hmn}), can be simplified as the follows: \(h_g = h_{0,0} = h_{1,1} = h_{4,4}=h_{5,5}\), these terms represent the energy of the system which both atoms are in ground energy level, and \(h_e = h_{10,10} = h_{11,11} = h_{14,4}=h_{15,15}\) represent the energy of the system which both atoms are in the excited energy level. Similarly, \(h_0 = (h_e+h_g)/2 = h_{2,2} = h_{3,3} = h_{6,6} = h_{7,7} = h_{8,8} = h_{9,9} = h_{12,12} = h_{13,13}\) correspond to the situation which one atom being excited and the other atom being in the ground state. We assume that the interaction does not alter the energy structure of the ground and excited levels. We can assign value \(h_e = -\hbar \omega/2\), \(h_g = \hbar \omega/2\), and \(h_0 = 0\) for simplicity. 

The entries, \((\hat{V}_{AB})_{mn} = \langle \alpha_i\beta_j \vert \hat{V}_{AB} \vert \alpha_{k} \beta_l\rangle\), correspond to the couplings in which the first atom's state change from \(\vert \alpha_k\rangle\) to \(\vert \alpha_i\rangle\) and the second atom's state change from \(\vert \beta_l\rangle\) to \(\vert\beta_j\rangle\). Since the interaction term \(\hat{V}_{AB}\) represents a two-atom interaction, it is reasonable to assume the entries correspond to the coupling in which only one atom state change would be zero. As an example, \((\hat{V}_{AB})_{10} = \langle \alpha_0\beta_1 \vert \hat{V}_{AB} \vert \alpha_{0} \beta_0\rangle = 0\) since only the state of the second atom changes. Second, we assume no coupling mechanism exists between the degenerate states within the same energy level. As an example, \((\hat{V}_{AB})_{41} = \langle \alpha_1\beta_0 \vert \hat{V}_{AB} \vert \alpha_{0} \beta_1\rangle = 0\) where both atoms remain in the degenerate ground energy level. Third, we assumed that only linearly polarized light couples to the atoms, like in the single-atom case. Any transition mediated by \(\pm 1\) photons is forbidden. This means terms like \((\hat{V}_{AB})_{9,2} = \langle \alpha_2\beta_1 \vert \hat{V}_{AB} \vert \alpha_{0} \beta_2\rangle = 0\). Lastly, by definition, we can write \((\hat{V}_{AB})_{ij} = (\hat{V}_{AB})_{ji}^*\).

Changing to the interaction picture, then using the rotating wave approximation. The interaction matrix has nonzero elements \(h_{2,8} = \langle \alpha_0 \beta_2 \vert \hat{V}_{AB} \vert \alpha_2 \beta_0 \rangle\), \(h_{3,9} = \langle \alpha_0 \beta_3 \vert \hat{V}_{AB} \vert \alpha_2 \beta_1\rangle\), \(h_{6,12} = \langle \alpha_1 \beta_2 \vert \hat{V}_{AB} \vert \alpha_3 \beta_0\rangle\), and \(h_{7,13} = \langle \alpha_1 \beta_3 \vert \hat{V}_{AB} \vert \alpha_3 \beta_1\rangle\). Observe that all the above terms describes the case which the first atom transition from the ground to the excited energy level while the second atom decays from the excited energy level to the ground energy level, it is reasonable to assume \(h = h_{2,8} =h_{3,9} =h_{6,12}=h_{7,13}\). We can write \(\hat{V}_{AB}^{(int)}\) as:
\begin{widetext}
\begin{equation}
    \hat{V}_{AB}^{(int)} = \left(
\begin{array}{cccccccccccccccc}
 0  & 0 & 0 & 0 & 0 & 0 & 0 & 0 & 0 & 0 & 0 & 0 & 0 & 0 & 0 & 0 \\
 0 & 0  & 0 & 0 & 0 & 0 & 0 & 0 & 0 & 0 & 0 & 0 & 0 & 0 & 0 & 0 \\
 0 & 0 & 0  & 0 & 0 & 0 & 0 & 0 & h & 0 & 0 & 0 & 0 & 0 & 0 & 0 \\
 0 & 0 & 0 & 0  & 0 & 0 & 0 & 0 & 0 & h & 0 & 0 & 0 & 0 & 0 & 0 \\
 0 & 0 & 0 & 0 & 0 & 0 & 0 & 0 & 0 & 0 & 0 & 0 & 0 & 0 & 0 & 0 \\
 0 & 0 & 0 & 0 & 0 & 0  & 0 & 0 & 0 & 0 & 0 & 0 & 0 & 0 & 0 & 0 \\
 0 & 0 & 0 & 0 & 0 & 0 & 0  & 0 & 0 & 0 & 0 & 0 & h & 0 & 0 & 0 \\
 0 & 0 & 0 & 0 & 0 & 0 & 0 & 0  & 0 & 0 & 0 & 0 & 0 & h & 0 & 0 \\
 0 & 0 & h^* & 0 & 0 & 0 & 0 & 0 & 0  & 0 & 0 & 0 & 0 & 0 & 0 & 0 \\
 0 & 0 & 0 & h^* & 0 & 0 & 0 & 0 & 0 & 0  & 0 & 0 & 0 & 0 & 0 & 0 \\
 0 & 0 & 0 & 0 & 0 & 0 & 0 & 0 & 0 & 0 & 0 & 0 & 0 & 0 & 0 & 0 \\
 0 & 0 & 0 & 0 & 0 & 0 & 0 & 0 & 0 & 0 & 0 & 0  & 0 & 0 & 0 & 0 \\
 0 & 0 & 0 & 0 & 0 & 0 & h^* & 0 & 0 & 0 & 0 & 0 & 0  & 0 & 0 & 0 \\
 0 & 0 & 0 & 0 & 0 & 0 & 0 & h^* & 0 & 0 & 0 & 0 & 0 & 0  & 0 & 0 \\
 0 & 0 & 0 & 0 & 0 & 0 & 0 & 0 & 0 & 0 & 0 & 0 & 0 & 0 & 0  & 0 \\
 0 & 0 & 0 & 0 & 0 & 0 & 0 & 0 & 0 & 0 & 0 & 0 & 0 & 0 & 0 & 0  \\
\end{array}
\right).
\end{equation}
\end{widetext}
The time evolution operator in the interaction picture is:
\begin{equation}
    \hat{U}_{AB}^{(int)}(t) = \exp\{- i \hat{V}_{AB}^{(int)} t/\hbar\}=\sum_{n=0}^{\infty} ( \frac{- i t}{\hbar})^n (\hat{V}_{AB}^{(int)})^{n}.
\end{equation}
Expanding the above time evolution operator explicitly, we would see that the time evolution takes the form given by Eq. (\ref{two atom unitary}) with a=0. Following the same procedure of Ref. \cite{AQJ}, we can use the above time evolution operator to construct a CZ gate. The procedure is shown in Appendix C, and the degenerate CZ gate takes the form:
\begin{widetext}
    \begin{equation}
    \hat{U}_5 = \left(
\begin{array}{cccccccccccccccc}
 1 & 0 & 0 & 0 & 0 & 0 & 0 & 0 & 0 & 0 & 0 & 0 & 0 & 0 & 0 & 0 \\
 0 & 1 & 0 & 0 & 0 & 0 & 0 & 0 & 0 & 0 & 0 & 0 & 0 & 0 & 0 & 0 \\
 0 & 0 & 1 & 0 & 0 & 0 & 0 & 0 & 0 & 0 & 0 & 0 & 0 & 0 & 0 & 0 \\
 0 & 0 & 0 & 1 & 0 & 0 & 0 & 0 & 0 & 0 & 0 & 0 & 0 & 0 & 0 & 0 \\
 0 & 0 & 0 & 0 & 1 & 0 & 0 & 0 & 0 & 0 & 0 & 0 & 0 & 0 & 0 & 0 \\
 0 & 0 & 0 & 0 & 0 & 1 & 0 & 0 & 0 & 0 & 0 & 0 & 0 & 0 & 0 & 0 \\
 0 & 0 & 0 & 0 & 0 & 0 & 1 & 0 & 0 & 0 & 0 & 0 & 0 & 0 & 0 & 0 \\
 0 & 0 & 0 & 0 & 0 & 0 & 0 & 1 & 0 & 0 & 0 & 0 & 0 & 0 & 0 & 0 \\
 0 & 0 & 0 & 0 & 0 & 0 & 0 & 0 & 1 & 0 & 0 & 0 & 0 & 0 & 0 & 0 \\
 0 & 0 & 0 & 0 & 0 & 0 & 0 & 0 & 0 & 1 & 0 & 0 & 0 & 0 & 0 & 0 \\
 0 & 0 & 0 & 0 & 0 & 0 & 0 & 0 & 0 & 0 & e^{4 i \theta} & 0 & 0 & 0 & 0 & 0 \\
 0 & 0 & 0 & 0 & 0 & 0 & 0 & 0 & 0 & 0 & 0 & e^{4 i \theta} & 0 & 0 & 0 & 0 \\
 0 & 0 & 0 & 0 & 0 & 0 & 0 & 0 & 0 & 0 & 0 & 0 & 1 & 0 & 0 & 0 \\
 0 & 0 & 0 & 0 & 0 & 0 & 0 & 0 & 0 & 0 & 0 & 0 & 0 & 1 & 0 & 0 \\
 0 & 0 & 0 & 0 & 0 & 0 & 0 & 0 & 0 & 0 & 0 & 0 & 0 & 0 & e^{4 i \theta} & 0 \\
 0 & 0 & 0 & 0 & 0 & 0 & 0 & 0 & 0 & 0 & 0 & 0 & 0 & 0 & 0 & e^{4 i \theta} \\
\end{array}
\right).
\end{equation}
\end{widetext}

Lastly, we note that \(U_5\) can be viewed as four separate CZ gates coupling different pairs of degenerate states. For example, the first, third, ninth, and eleventh diagonal entries form a CZ gate between the degenerate ground state \(\vert\alpha_0\rangle, \vert\beta_0\rangle\), and the degenerate excited state \(\vert \alpha_2 \rangle, \vert \beta_2\rangle\). Similarly, the second, fourth, tenth, and twelfth entry forms the CZ gate between the degenerate ground state \(\vert\alpha_0\rangle, \vert\beta_1\rangle\) and the degenerate excited state \(\vert \alpha_2 \rangle, \vert \beta_3\rangle\), and so on. Therefore, under the assumption of identical atoms, the same interaction strength, and linearly polarized light-mediated coupling, it is postulated that performing the CZ gate is plausible.

\section{Conclusion}
Due to the spherical symmetry of the atom, there is no fundamental law that forbids the construction of a trapped-ion-based quantum computer with degenerate levels. An ideal degenerate Hadamard gate can be constructed with \(^2S_{1/2}\) and \(^2P_{1/2}\) fine structure states with no external static magnetic field present. The two-level degenerate Hilbert space formed by all \(^2S_{1/2}\) and \(^2P_{1/2}\) fine structure states can be decomposed into superpositions of identical rank-two subspaces, and the transition operations can be identically applied to each subspace by defining the quantization axis of the atom to be the polarization axis of the electric field. Using the above result, we can construct the degenerate Hadamard gate.  

The presence of a weak static magnetic field satisfying \(g_s\mu_B\vert \boldsymbol{B}\vert/\hbar\Omega\ll 1\) allows us to expand the time evolution operator as a power series of the magnetic field coupling strength over the electric dipole coupling strength. With the assumption that the Rabi frequency has a value greater than \(10^3Hz\), such a bound requires the magnetic field to be much smaller than \(10^{-8}\)T. With the current \(\mu\) metal shielding, such a bound is attainable. By performing an expansion with the above constraint on the environmental magnetic field, we showed that the zeroth-order expansion recovers the perfect degenerate Hadamard gate, while higher orders act as corrections. The fidelity of the degenerate Hadamard gate up to the second order is given in Eq. (\ref{fidelity}). The first nontrivial correction is of the order \((\mu_Bg_s\vert \boldsymbol{B}\vert /\hbar\Omega)^2\). 

We also discussed the conditions required to perform a Controlled-Z (CZ) gate. Again, assuming the atoms are identical, a controlled Z gate can be constructed from a time evolution operator of two atoms (\ref{two atom unitary}). The validity of the CZ gate is based on the assumption of identical atoms, the same interaction strength, and linearly polarized light-mediated coupling and no single-atom internal state transfer.

Therefore, we conclude that degeneracy lifting is not necessary for a trapped ion qubit with two levels separated by a single frequency. Using degenerate states could reduce experimental difficulty in providing a stable Zeeman field and increase the flexibility of the encoding strategy. However, as a cautious note, one should not attempt to draw an analogy between a degenerate atom and a hyperentangled system \cite{Kwiat} since the resemblance is only superficial. The total angular momentum quantum number (J) and the secondary total angular momentum (\(M_j\)) cannot be operated independently without considering the ancillary states; hence, we do not consider the two degrees of freedom to be truly hyperentangled. Similarly, the degeneracy does not provide an extra capability for error correction, as these techniques rely on the presence of ingenious ancilla qubits that enable us to perform syndrome measurements without disturbing the superposition. Finally, the connection between the embedded symmetry of the system and its degeneracy is well known \cite{Elliott}. Therefore, we proposed that there might be a general relation between the symmetry and the system's qubit or qudit-like behaviours, and we left the topic for future research.

\appendix
\begin{widetext}
\section{An Example for Hadamard Gate with Degenerate Levels}
As a concrete example, let's compute the transition matrix for the \(^2S_{1/2}\) and \(^2P_{1/2}\) levels. Following the previous notation, the two \(^2S_{1/2}\) degenerate states are \(\vert 0, m_0=-\frac{1}{2}\rangle\) and \(\vert 0, m_1=\frac{1}{2}\rangle\). The corresponding coefficients are \(\alpha_0(t)\) and \(\alpha_1(t)\). The two \(^2P_{1/2}\) degenerate states are \(\vert 1, m_0=-\frac{1}{2}\rangle\) and \(\vert 1, m_1=\frac{1}{2}\rangle\) and has corresponding coefficients \(\beta_0(t)\) and \(\beta_1(t)\). Using the symmetry argument and Eq. (\ref{omega value}), we find \(\Omega_{m_im_j} = 1/\sqrt{6} \delta_{m_im_j}\). We define a new parameter representing the transition strength: \(\Omega=\frac{\vert \boldsymbol{E}\vert  S}{\hbar \sqrt{6}}\). The matrix equation is simplified to:
\begin{equation}
    \frac{d}{dt}\begin{pmatrix}
        \alpha_0 \\
        \alpha_1 \\
        \beta_0 \\
        \beta_1
    \end{pmatrix}=-i\frac{\Omega}{2}\begin{pmatrix}
        0 & 0 & 1 & 0\\
        0 & 0 & 0 & 1 \\
        1 & 0 & 0 &0 \\
        0 & 1 & 0 &0
    \end{pmatrix} \begin{pmatrix}
        \alpha_0 \\
        \alpha_1 \\
        \beta_0 \\
        \beta_1
    \end{pmatrix}.
\end{equation}
Solving the above differential equation with the initial condition \(\vert \psi(0) \rangle = \alpha_0(0) \vert 0, m_0=-\frac{1}{2} \rangle + \alpha_1(0) \vert 0, m_1=\frac{1}{2}\rangle\), the solution is:
\begin{equation}
\begin{split}
    \alpha_0(t) & = \alpha_0(0) \cos{\frac{\Omega t}{2}}, \\
    \alpha_1(t) & = \alpha_1(0) \cos{\frac{\Omega t}{2}}, \\
    \beta_0(t) & = -i \alpha_0(0) \sin{\frac{\Omega t}{2}}, \\
    \beta_1(t) & = -i \alpha_1(0) \sin{\frac{\Omega t}{2}}.
\end{split}
\end{equation}

Therefore, the state as a function of time is:

\begin{equation}\label{result1}
\begin{split}
    \vert \tilde{\psi}(t)\rangle & = \alpha_0(0) [\cos{\frac{\Omega t}{2}} \vert 0,0\rangle -i \sin{\frac{\Omega t}{2}} \vert 1,0\rangle] \\
    & + \alpha_1(0) [\cos{\frac{\Omega t}{2}} \vert 0, 1\rangle - i \sin{\frac{\Omega t}{2}} \vert 1, 1\rangle].
\end{split}
\end{equation}
The equation above shows that the Rabi oscillation occurs independently in the subspace of the degenerate state pair \(\vert 0,m_0=-\frac{1}{2}\rangle\), \(\vert 1,m_0=\frac{1}{2} \rangle\) and the pair \(\vert 0,m_0=\frac{1}{2} \rangle\), \(\vert 1,m_0=\frac{1}{2}\rangle\). The existence of the Rabi oscillation means that there should be no difference between whether or not we choose to lift degeneracy.

Another way to view the degenerate pairs acting like qubits is to write down the time evolution operator and use it to build a Hadamard gate. We already computed that the effective Hamiltonian in the interaction picture:
\begin{equation}
    \hat{\mathcal{H}}_{int}=\frac{\hbar\Omega}{2}\begin{pmatrix}
        0&0&1&0\\
        0&0&0&1\\
        1&0&0&0\\
        0&1&0&0
    \end{pmatrix}.
\end{equation}

The time evolution in the interaction picture is then:

\begin{equation}
    \hat{U}_{int}(t) = \begin{pmatrix}
        \cos(\frac{\Omega t}{2})& 0& -i \sin(\frac{\Omega t}{2})& 
  0\\ 
  0& \cos(\frac{\Omega t}{2})& 
  0& -i \sin(\frac{\Omega t}{2}) \\
  -i \sin(\frac{\Omega t}{2})&
   0& \cos(\frac{\Omega t}{2})& 
  0 \\
  0& -i \sin(\frac{\Omega t}{2})& 0& 
  \cos(\frac{\Omega t}{2}) 
    \end{pmatrix}.
\end{equation}
Define the free evolution of the state:
\begin{equation}
    \hat{U}_0(t) = \exp[- i \hat{\mathcal{H}}_0t/\hbar] = \begin{pmatrix}
        \exp[it\omega/2]&0&0&0\\
        0&\exp[it\omega/2]&0&0\\
        0&0&\exp[-it\omega/2]&0\\
        0&0&0&\exp[-it\omega/2]
    \end{pmatrix},
\end{equation}
where \(\omega=(E_{^2P_{1/2}}-E_{^2S_{1/2}})/\hbar\). Since the states described by the interaction picture \(\vert \tilde{\psi}(t)\rangle\) and the states described by the Schrodinger picture are connected by:
\begin{equation}
    \vert\psi(t)\rangle=U_0(t)\vert \tilde{\psi}(t)\rangle,
\end{equation}
The effective time evolution in the Schrodinger picture is given by:
\begin{equation}\label{U(t)}
    \hat{U}(t)=\hat{U}_0(t)\hat{U}_{int}(t)=\begin{pmatrix}
        e^{i t \omega/2} \cos \left(\frac{\Omega t}{2  }\right) & 0 & -i e^{i t \omega/2} \sin \left(\frac{\Omega t}{2}\right) & 0 \\
 0 & e^{i t \omega/2} \cos \left(\frac{\Omega t}{2}\right) & 0 & -i e^{i t \omega/2} \sin \left(\frac{\Omega t}{2}\right) \\
 -i e^{-i t \omega/2} \sin \left(\frac{\Omega t}{2 }\right) & 0 & e^{-i t \omega/2} \cos \left(\frac{\Omega t}{2}\right) & 0 \\
 0 & -i e^{-i t \omega/2} \sin \left(\frac{\Omega t}{2}\right) & 0 & e^{-i t \omega/2} \cos \left(\frac{\Omega t}{2}\right) \\
    \end{pmatrix}.
\end{equation}
We can build the degenerate Hadamard gate with a series of time evolution: \( \hat{U}_0(\frac{3\pi}{2\omega})\hat{U}(\frac{ \pi}{2 \Omega})\hat{U}_0(\frac{3\pi}{2\omega})\). Each pair of degenerate states with the same index undergoes identical operations under our degenerate Hadamard gate:
\begin{equation}
    \hat{U}_0(\frac{3\pi}{2\omega})\hat{U}(\frac{ \pi}{2 \Omega})\hat{U}_0(\frac{3\pi}{2\omega}) = -i \frac{1}{\sqrt{2}}\begin{pmatrix}
    1&0&1&0\\
    0&1&0&1\\
    1&0&-1&0\\
    0&1&0&-1\\
    \end{pmatrix}.
\end{equation}

Now we see that the degenerate Hadamard gate is just regular Hadamard gates applied to each of the qubits formed by \(\alpha_0(t), \beta_0(t)\)  and \(\alpha_1(t), \beta_1(t)\). However, no entanglement can exist since the two qubits are in superposition in Eq. (\ref{result1}).
More generally, any operation applied to the Hilbert space of degenerate states can be viewed as two identical operations, each applied to the two rank-2 subspaces formed by the pair \(\alpha_0,\beta_0\) and the pair \(\alpha_1,\beta_1\). Each of the subspaces resembles a qubit, and the total degenerate state is a superposition of these qubit subspaces.

\section{Fidelity of Hadamard Gate}
Let us follow the proof given by \cite{Pedersen} and show the equation holds by explicitly computing the left-hand side of the equation using the integration technique from \cite{Ambainis}. For simplicity, define \(\hat{M}=(\hat{H})(\hat{U}_{Had})\). We write the linear operator as a sum of its Hermitian part \(\hat{M}_s\) and its anti-Hermitian part \(\hat{M}_a\):
\begin{equation}
    \hat{M} = \frac{\hat{M}+\hat{M}^{\dag}}{2}+\frac{\hat{M}-\hat{M}^{\dag}}{2} =\hat{M}_s+\hat{M}_a
\end{equation}
By proving Eq. (\ref{fidelity}) holds for \(M_s\) and \(M_a\) separately, we can show that Eq. (\ref{fidelity}) holds for arbitrary M.
Since \(M_s\) is Hermitian, it can be diagonalized by a unitary matrix \(\chi\), and so we can write:
\begin{equation}
    I =\int_{S^{2n-1}}\vert\langle \psi\vert \hat{M}_s\vert\psi\rangle\vert^2 d\sigma/V_{S^{2n-1}} = \int_{S^{2n-1}}\vert(\langle \psi\vert \chi^{\dag}) \hat{\Lambda} (\chi\vert\psi\rangle)\vert^2 d\sigma/V_{S^{2n-1}}.
\end{equation}
By defining \(\vert \phi\rangle=\chi\vert\psi\rangle\), the equation becomes:
\begin{equation}
    I = \int_{S^{2n-1}}\vert\langle \phi\vert  \hat{\Lambda} \vert\phi\rangle\vert^2 d\sigma/V_{S^{2n-1}}.
\end{equation}
The integration boundary remains unchanged since we are integrating over all states located on the \(S^{2n-1}\) unit sphere. Now, expand \(\vert\phi\rangle\) with a chosen basis, \(\vert \phi\rangle = \sum_{i}c_i\vert i\rangle\), the equation becomes:
\begin{equation}
    I = \sum_{ij}\Lambda_{i}\Lambda_{j}\int_{S^{2n-1}}\vert c_i\vert^2 \vert c_j\vert^2 d\sigma/V_{S^{2n-1}}.
\end{equation}
Now, instead of integrating over the surface of a sphere, we can integrate all space \(R^{2n}\) by inserting a delta function inside the integral to make the value of the function effectively zero outside of the unit shell.
\begin{equation}
    I=\sum_{ij}\Lambda_{i}\Lambda_{j}\int_{R^{2n}}(\vert x_i\vert^2+\vert y_i\vert^2)( \vert x_j\vert^2 +\vert y_j\vert^2 ) \delta((\sum_p \vert x_p\vert^2+\vert y_p\vert^2)-1) \prod_kdx_kdy_k/V_{S^{2n-1}}.
\end{equation}
Perform a change of variable, let \(u_i/r = x_i\) and \(v_i/r=y_i\), we obtain:
\begin{equation}
\begin{split}
    I&=\sum_{ij}\Lambda_{i}\Lambda_{j}\int_{R^{2n}}r^{-4}(\vert u_i\vert^2+\vert v_i\vert^2)( \vert u_j\vert^2 +\vert v_j\vert^2 ) \delta(\frac{\sqrt{\sum_p \vert u_p\vert^2+\vert v_p\vert^2}}{r}-1) r^{-2n}\prod_kdu_kdv_k/V_{S^{2n-1}}\\
    & =\sum_{ij}\Lambda_{i}\Lambda_{j}\int_{R^{2n}}r^{-4}(\vert u_i\vert^2+\vert v_i\vert^2)( \vert u_j\vert^2 +\vert v_j\vert^2 ) r\delta(\sqrt{\sum_p \vert u_p\vert^2+\vert v_p\vert^2}-r) r^{-2n}\prod_kdu_kdv_k/V_{S^{2n-1}},
\end{split}
\end{equation}
where we used the property, \(\delta(a/b-1)=b\delta(a-b)\). Multiply both sides by \(r^{2n+4-1}e^{-r^2}\), then integrate with respect to r from 0 to infinity:
\begin{equation}
    I\int_{0}^{\infty} r^{2(n+2)-1}e^{-r^2}dr = \sum_{ij}\Lambda_{i}\Lambda_{j}\int_0^{\infty}\int_{R^{2n}}(\vert u_i\vert^2+\vert v_i\vert^2)( \vert u_j\vert^2 +\vert v_j\vert^2 ) \delta(\sqrt{\sum_p \vert u_p\vert^2+\vert v_p\vert^2}-r) \prod_kdu_kdv_k e^{-r^2}dr/V_{S^{2n-1}},
\end{equation}
and exchange the order of integration, so we integrate with respect to r first, we obtain:
\begin{equation}
    I\frac{\Gamma(n+2)}{2} = \sum_{ij}\Lambda_{i}\Lambda_{j}\int_{R^{2n}}(\vert u_i\vert^2+\vert v_i\vert^2)( \vert u_j\vert^2 +\vert v_j\vert^2 )  e^{-(\sum_p\vert u_p\vert^2+\vert v_p\vert^2)}\prod_kdu_kdv_k /V_{S^{2n-1}}.
\end{equation}
\begin{equation}
\begin{split}
    I = \frac{2}{(n+1)!} [& \sum_i \Lambda_{i}^2\int_{R^{2n}}(\vert u_i\vert^2+\vert v_i\vert^2)^2  e^{-(\sum_p\vert u_p\vert^2+\vert v_p\vert^2)}\prod_kdu_kdv_k \\
    & + \sum_{i\neq j}\Lambda_{i}\Lambda_{j}\int_{R^{2n}}(\vert u_i\vert^2+\vert v_i\vert^2)( \vert u_j\vert^2 +\vert v_j\vert^2 )  e^{-(\sum_p\vert u_p\vert^2+\vert v_p\vert^2)}\prod_kdu_kdv_k ]/V_{S^{2n-1}}.
\end{split}
\end{equation}
Perform a change of variable again, let \(\vert u_i\vert^2+\vert v_i\vert^2 = r^2\), the first integral becomes:
\begin{equation}
\begin{split}
    \int_{R^{2n}}(\vert u_i\vert^2+\vert v_i\vert^2)^2  e^{-(\sum_p\vert u_p\vert^2+\vert v_p\vert^2)}\prod_kdu_kdv_k& = \int_{R^2} r^4 e^{-r^2} rdr d\theta(\int_{R^2}e^{-r^2}rdrd\theta)^{n-1}\\
    & = 2\pi^n.
\end{split}
\end{equation}
The second integral gives:
\begin{equation}
\begin{split}
     &\int_{R^{2n}}(\vert u_i\vert^2+\vert v_i\vert^2)(\vert u_j\vert^2+\vert v_j\vert^2)e^{-(\sum_p\vert u_p\vert^2+\vert v_p\vert^2)}\prod_k du_kdv_k \\
     & = (\int_{R^2}(\vert u_i\vert^2 + \vert v_i\vert^2) e^{-(\vert u_i\vert^2+\vert v_i\vert^2)}du_idv_i)^2(\int_{R^2} e^{-(\vert u_p\vert^2+\vert v_p\vert^2)}du_pdv_p)^{n-2}\\
     & = \pi^n.
\end{split}
\end{equation}
Therefore, we find:
\begin{equation}
    I = \frac{2}{(n+1)! V_{S^{2n-1}}}[2\pi^n\sum_i\Lambda_i^2+\pi^n\sum_{i\neq j}\Lambda_i\Lambda_j].
\end{equation}
Substitute the surface area of the unit sphere in 2n dimension, \(V_{S^{2n-1}} = 2\pi^2/(n-1)!\), the equation equals to:
\begin{equation}
    I = \frac{2}{(n+1)!}\frac{(n-1)!}{2\pi^n}[2\pi^n\sum_i\Lambda_i^2+\pi^n\sum_{i\neq j}\Lambda_i\Lambda_j] =\frac{1}{n(n+1)}[\sum_i\Lambda_i^2 + (\sum_i\Lambda_i^2 + \sum_{i\neq j}\Lambda_i\Lambda_j)].
\end{equation}
Now write \(Tr(\hat{\Lambda}^2) =\sum_i\Lambda_i^2\), and \(Tr(\Lambda)^2 = \sum_{i}\Lambda_i^2 + \sum_{i\neq j}\Lambda_i\Lambda_j\), since all \(\Lambda_i\) are real, and traces are basis independent, Eq. (\ref{fidelity}) holds for Hermitian M. For the anti-Hermitian case, write the anti-Hermitian matrix as a Hermitian matrix multiply by i. A similar analysis would show that Eq. (\ref{fidelity}) also holds. Therefore, we conclude that Eq. (\ref{fidelity}) is valid.

\section{Controlled Z Gate between two Qubits}
The power of quantum computation lies in the ability to perform two-qubit gates. Here, we generalized the approach given in \cite{AQJ} to degenerate systems. Using the same strategy presented in the reference, we attempt to set limitations on the interaction Hamiltonian such that the time evolution of the combined system takes the form:
\begin{equation}
\label{two atom unitary}
    \hat{U}(t) = \left(
\begin{array}{cccccccccccccccc}
 e^{i a t} & 0 & 0 & 0 & 0 & 0 & 0 & 0 & 0 & 0 & 0 & 0 & 0 & 0 & 0 & 0 \\
 0 & e^{i a t} & 0 & 0 & 0 & 0 & 0 & 0 & 0 & 0 & 0 & 0 & 0 & 0 & 0 & 0 \\
 0 & 0 & c^* & 0 & 0 & 0 & 0 & 0 & d & 0 & 0 & 0 & 0 & 0 & 0 & 0 \\
 0 & 0 & 0 & c^* & 0 & 0 & 0 & 0 & 0 & d & 0 & 0 & 0 & 0 & 0 & 0 \\
 0 & 0 & 0 & 0 & e^{i a t} & 0 & 0 & 0 & 0 & 0 & 0 & 0 & 0 & 0 & 0 & 0 \\
 0 & 0 & 0 & 0 & 0 & e^{i a t} & 0 & 0 & 0 & 0 & 0 & 0 & 0 & 0 & 0 & 0 \\
 0 & 0 & 0 & 0 & 0 & 0 & c^* & 0 & 0 & 0 & 0 & 0 & d & 0 & 0 & 0 \\
 0 & 0 & 0 & 0 & 0 & 0 & 0 & c^* & 0 & 0 & 0 & 0 & 0 & d & 0 & 0 \\
 0 & 0 & -d^* & 0 & 0 & 0 & 0 & 0 & c & 0 & 0 & 0 & 0 & 0 & 0 & 0 \\
 0 & 0 & 0 & -d^* & 0 & 0 & 0 & 0 & 0 & c & 0 & 0 & 0 & 0 & 0 & 0 \\
 0 & 0 & 0 & 0 & 0 & 0 & 0 & 0 & 0 & 0 & e^{i a t} & 0 & 0 & 0 & 0 & 0 \\
 0 & 0 & 0 & 0 & 0 & 0 & 0 & 0 & 0 & 0 & 0 & e^{i a t} & 0 & 0 & 0 & 0 \\
 0 & 0 & 0 & 0 & 0 & 0 & -d^* & 0 & 0 & 0 & 0 & 0 & c & 0 & 0 & 0 \\
 0 & 0 & 0 & 0 & 0 & 0 & 0 & -d^* & 0 & 0 & 0 & 0 & 0 & c & 0 & 0 \\
 0 & 0 & 0 & 0 & 0 & 0 & 0 & 0 & 0 & 0 & 0 & 0 & 0 & 0 & e^{i a t} & 0 \\
 0 & 0 & 0 & 0 & 0 & 0 & 0 & 0 & 0 & 0 & 0 & 0 & 0 & 0 & 0 & e^{i a t} \\
\end{array}
\right),
\end{equation}
where c and h are arbitrary complex functions of time. We examine the assumptions to produce the above time evolution and show that this unitary evolution is sufficient to construct a CZ gate.

We start by considering the interaction between two atoms, each degenerate in the \(^2S_{1/2}\) ground and \(^2P_{1/2}\) excited levels. We label the basis for the first atom to be \(\vert \alpha_{i}\rangle\), and for the second atom, \(\vert \beta_{j}\rangle\), where \(i,j=0,1,2,3\) correspond to the ground and the excited energy level with degeneracy. Then, the combined basis is \(\vert \alpha_{i}\beta_{j}\rangle\). Let \(\hat{\mathcal{H}}_{AB}\) be the Hamiltonian describe the two atoms system, then \(\hat{\mathcal{H}}_{AB}\) can be represented in the given basis \(\vert \alpha_{i}\beta_{j}\rangle\) with matrix elements given by:
\begin{equation}\label{hmn}
    h_{mn}= \langle \alpha_i \beta_j \vert \hat{\mathcal{H}} \vert \alpha_k \beta_l\rangle,
\end{equation}
with \(m = 4i+j, n=4k+l\). 
The matrix \(\hat{\mathcal{H}}_{AB}\) can be decomposed as:
\begin{equation}
    \hat{\mathcal{H}}_{AB} = \hat{\mathcal{H}}_{A}\otimes \hat{I} + \hat{I} \otimes \hat{\mathcal{H}}_{B}+\hat{V}_{AB},
\end{equation}
where \(\hat{I}\) is the identity matrix, \(\hat{\mathcal{H}}_{A}\) and \(\hat{\mathcal{H}}_{B}\) are matrices representing the individual Hamiltonian for the first and the second atom in the basis \(\vert\alpha_i\rangle, \vert\beta_j\rangle\) correspondingly, and \(\hat{V}_{AB}\) is the matrix represent the interaction between the two atoms.
In our chosen basis, assuming the two atoms are identical, \(\hat{\mathcal{H}}_A\) and \(\hat{\mathcal{H}}_B\) take the form:
\begin{equation}
    \hat{\mathcal{H}}_A = \hat{\mathcal{H}}_B = \frac{\hbar\omega}{2}\begin{pmatrix}
        1 & 0 &0&0\\
        0& 1 &0&0\\
        0&0&-1&0\\
        0&0&0&-1
    \end{pmatrix},
\end{equation}
with \(\omega = (E_g-E_e)/\hbar\). The interaction \(\hat{V}_{AB}\) is given by:
\begin{equation}
    \hat{V}_{AB} =\hat{\mathcal{H}}_{AB}-\hat{\mathcal{H}}_{A}\otimes \hat{I} - \hat{I} \otimes \hat{\mathcal{H}}_{B}.
\end{equation}
Explicitly, \((\hat{V}_{AB})_{ii} = h_{ii}-\hbar \omega/2\) for i = \{0,1,...,7\}, \((\hat{V}_{AB})_{ii} = h_{ii}+\hbar \omega/2\) for i = \{8,9,...,15\}, and \((\hat{V}_{AB})_{ij} = h_{ij}\) for \(i \neq j\). Now, we would like to consider the meaning of each interaction term and find the required assumption that Eq. (\ref{two atom unitary}) is true. 

Since we assume the two atoms are identical, the diagonal entries of the total Hamiltonian, Eq. (\ref{hmn}), can be simplified as the follows: \(h_g = h_{0,0} = h_{1,1} = h_{4,4}=h_{5,5}\), these terms represent the energy of the system which both atoms are in ground energy level, and \(h_e = h_{10,10} = h_{11,11} = h_{14,4}=h_{15,15}\) represent the energy of the system which both atoms are in the excited energy level. Similarly, \(h_0 = (h_e+h_g)/2 = h_{2,2} = h_{3,3} = h_{6,6} = h_{7,7} = h_{8,8} = h_{9,9} = h_{12,12} = h_{13,13}\) correspond to the situation which one atom being excited and the other atom being in the ground state. We assume that the interaction does not alter the energy structure of the ground and excited levels. We can assign value \(h_e = -\hbar \omega/2\), \(h_g = \hbar \omega/2\), and \(h_0 = 0\) for simplicity. 

The entries, \((\hat{V}_{AB})_{mn} = \langle \alpha_i\beta_j \vert \hat{V}_{AB} \vert \alpha_{k} \beta_l\rangle\), correspond to the couplings in which the first atom's state change from \(\vert \alpha_k\rangle\) to \(\vert \alpha_i\rangle\) and the second atom's state change from \(\vert \beta_l\rangle\) to \(\vert\beta_j\rangle\). Since the interaction term \(\hat{V}_{AB}\) represents a two-atom interaction, it is reasonable to assume the entries correspond to the coupling in which only one atom state change would be zero. As an example, \((\hat{V}_{AB})_{10} = \langle \alpha_0\beta_1 \vert \hat{V}_{AB} \vert \alpha_{0} \beta_0\rangle = 0\) since only the state of the second atom changes. Second, we assume no coupling mechanism exists between the degenerate states within the same energy level. As an example, \((\hat{V}_{AB})_{41} = \langle \alpha_1\beta_0 \vert \hat{V}_{AB} \vert \alpha_{0} \beta_1\rangle = 0\) where both atoms remain in the degenerate ground energy level. Third, we assumed that only linearly polarized light couples to the atoms, like in the single-atom case. Any transition mediated by \(\pm 1\) photons is forbidden. This means terms like \((\hat{V}_{AB})_{9,2} = \langle \alpha_2\beta_1 \vert \hat{V}_{AB} \vert \alpha_{0} \beta_2\rangle = 0\). Lastly, by definition, we can write \((\hat{V}_{AB})_{ij} = (\hat{V}_{AB})_{ji}^*\).

Changing to the interaction picture using the local evolution operator given by:
\begin{equation}
    \hat{U}_{local} = \exp\{i\hbar\omega t/2 (\hat{Z} \otimes \hat{I})\}\otimes\exp\{i\hbar\omega t/2 (\hat{Z} \otimes \hat{I})\},
\end{equation}
The interaction matrix \(\hat{V}_{AB}^{(int)}\) is given by:
\begin{equation}
    \hat{V}_{AB}^{(int)} = \hat{U}_{local} \hat{V}_{AB} \hat{U}_{local}^{\dag}.
\end{equation}
Then we use the rotating wave approximation to set the fast-rotating terms, proportional to \(\exp(\pm i\hbar\omega t/2)\) to zero. The interaction matrix is given by:
\begin{equation}
    \hat{V}_{AB}^{(int)} = \left(
\begin{array}{cccccccccccccccc}
 0  & 0 & 0 & 0 & 0 & 0 & 0 & 0 & 0 & 0 & 0 & 0 & 0 & 0 & 0 & 0 \\
 0 & 0  & 0 & 0 & 0 & 0 & 0 & 0 & 0 & 0 & 0 & 0 & 0 & 0 & 0 & 0 \\
 0 & 0 & 0  & 0 & 0 & 0 & 0 & 0 & h_{2,8} & 0 & 0 & 0 & 0 & 0 & 0 & 0 \\
 0 & 0 & 0 & 0  & 0 & 0 & 0 & 0 & 0 & h_{3,9} & 0 & 0 & 0 & 0 & 0 & 0 \\
 0 & 0 & 0 & 0 & 0 & 0 & 0 & 0 & 0 & 0 & 0 & 0 & 0 & 0 & 0 & 0 \\
 0 & 0 & 0 & 0 & 0 & 0  & 0 & 0 & 0 & 0 & 0 & 0 & 0 & 0 & 0 & 0 \\
 0 & 0 & 0 & 0 & 0 & 0 & 0  & 0 & 0 & 0 & 0 & 0 & h_{6,12} & 0 & 0 & 0 \\
 0 & 0 & 0 & 0 & 0 & 0 & 0 & 0  & 0 & 0 & 0 & 0 & 0 & h_{7,13} & 0 & 0 \\
 0 & 0 & h_{2,8}^* & 0 & 0 & 0 & 0 & 0 & 0  & 0 & 0 & 0 & 0 & 0 & 0 & 0 \\
 0 & 0 & 0 & h_{3,9}^* & 0 & 0 & 0 & 0 & 0 & 0  & 0 & 0 & 0 & 0 & 0 & 0 \\
 0 & 0 & 0 & 0 & 0 & 0 & 0 & 0 & 0 & 0 & 0 & 0 & 0 & 0 & 0 & 0 \\
 0 & 0 & 0 & 0 & 0 & 0 & 0 & 0 & 0 & 0 & 0 & 0  & 0 & 0 & 0 & 0 \\
 0 & 0 & 0 & 0 & 0 & 0 & h_{6,12}^* & 0 & 0 & 0 & 0 & 0 & 0  & 0 & 0 & 0 \\
 0 & 0 & 0 & 0 & 0 & 0 & 0 & h_{7,13}^* & 0 & 0 & 0 & 0 & 0 & 0  & 0 & 0 \\
 0 & 0 & 0 & 0 & 0 & 0 & 0 & 0 & 0 & 0 & 0 & 0 & 0 & 0 & 0  & 0 \\
 0 & 0 & 0 & 0 & 0 & 0 & 0 & 0 & 0 & 0 & 0 & 0 & 0 & 0 & 0 & 0  \\
\end{array}
\right).
\end{equation}
Observe that \(h_{2,8} = \langle \alpha_0 \beta_2 \vert \hat{V}_{AB} \vert \alpha_2 \beta_0 \rangle\), \(h_{3,9} = \langle \alpha_0 \beta_3 \vert \hat{V}_{AB} \vert \alpha_2 \beta_1\rangle\), \(h_{6,12} = \langle \alpha_1 \beta_2 \vert \hat{V}_{AB} \vert \alpha_3 \beta_0\rangle\), and \(h_{7,13} = \langle \alpha_1 \beta_3 \vert \hat{V}_{AB} \vert \alpha_3 \beta_1\rangle\) all describes the case which the first atom transition from the ground to the excited energy level while the second atom decays from the excited energy level to the ground energy level, it is reasonable to assume \(h = h_{2,8} =h_{3,9} =h_{6,12}=h_{7,13}\). We can write \(\hat{V}_{AB}^{(int)}\) as:
\begin{equation}
    \hat{V}_{AB}^{(int)} = \left(
\begin{array}{cccccccccccccccc}
 0  & 0 & 0 & 0 & 0 & 0 & 0 & 0 & 0 & 0 & 0 & 0 & 0 & 0 & 0 & 0 \\
 0 & 0  & 0 & 0 & 0 & 0 & 0 & 0 & 0 & 0 & 0 & 0 & 0 & 0 & 0 & 0 \\
 0 & 0 & 0  & 0 & 0 & 0 & 0 & 0 & h & 0 & 0 & 0 & 0 & 0 & 0 & 0 \\
 0 & 0 & 0 & 0  & 0 & 0 & 0 & 0 & 0 & h & 0 & 0 & 0 & 0 & 0 & 0 \\
 0 & 0 & 0 & 0 & 0 & 0 & 0 & 0 & 0 & 0 & 0 & 0 & 0 & 0 & 0 & 0 \\
 0 & 0 & 0 & 0 & 0 & 0  & 0 & 0 & 0 & 0 & 0 & 0 & 0 & 0 & 0 & 0 \\
 0 & 0 & 0 & 0 & 0 & 0 & 0  & 0 & 0 & 0 & 0 & 0 & h & 0 & 0 & 0 \\
 0 & 0 & 0 & 0 & 0 & 0 & 0 & 0  & 0 & 0 & 0 & 0 & 0 & h & 0 & 0 \\
 0 & 0 & h^* & 0 & 0 & 0 & 0 & 0 & 0  & 0 & 0 & 0 & 0 & 0 & 0 & 0 \\
 0 & 0 & 0 & h^* & 0 & 0 & 0 & 0 & 0 & 0  & 0 & 0 & 0 & 0 & 0 & 0 \\
 0 & 0 & 0 & 0 & 0 & 0 & 0 & 0 & 0 & 0 & 0 & 0 & 0 & 0 & 0 & 0 \\
 0 & 0 & 0 & 0 & 0 & 0 & 0 & 0 & 0 & 0 & 0 & 0  & 0 & 0 & 0 & 0 \\
 0 & 0 & 0 & 0 & 0 & 0 & h^* & 0 & 0 & 0 & 0 & 0 & 0  & 0 & 0 & 0 \\
 0 & 0 & 0 & 0 & 0 & 0 & 0 & h^* & 0 & 0 & 0 & 0 & 0 & 0  & 0 & 0 \\
 0 & 0 & 0 & 0 & 0 & 0 & 0 & 0 & 0 & 0 & 0 & 0 & 0 & 0 & 0  & 0 \\
 0 & 0 & 0 & 0 & 0 & 0 & 0 & 0 & 0 & 0 & 0 & 0 & 0 & 0 & 0 & 0  \\
\end{array}
\right).
\end{equation}
The time evolution operator in the interaction picture is:
\begin{equation}
    \hat{U}_{AB}^{(int)}(t) = \exp\{- i \hat{V}_{AB}^{(int)} t/\hbar\}=\sum_{n=0}^{\infty} ( \frac{- i t}{\hbar})^n (\hat{V}_{AB}^{(int)})^{n}.
\end{equation}
Define a matrix \(\hat{D}\):
\begin{equation}
    \hat{D} =
     \left(
\begin{array}{cccccccccccccccc}
 0  & 0 & 0 & 0 & 0 & 0 & 0 & 0 & 0 & 0 & 0 & 0 & 0 & 0 & 0 & 0 \\
 0 & 0  & 0 & 0 & 0 & 0 & 0 & 0 & 0 & 0 & 0 & 0 & 0 & 0 & 0 & 0 \\
 0 & 0 & 1  & 0 & 0 & 0 & 0 & 0 & 0 & 0 & 0 & 0 & 0 & 0 & 0 & 0 \\
 0 & 0 & 0 & 1  & 0 & 0 & 0 & 0 & 0 & 0 & 0 & 0 & 0 & 0 & 0 & 0 \\
 0 & 0 & 0 & 0 & 0 & 0 & 0 & 0 & 0 & 0 & 0 & 0 & 0 & 0 & 0 & 0 \\
 0 & 0 & 0 & 0 & 0 & 0  & 0 & 0 & 0 & 0 & 0 & 0 & 0 & 0 & 0 & 0 \\
 0 & 0 & 0 & 0 & 0 & 0 & 1  & 0 & 0 & 0 & 0 & 0 & 0 & 0 & 0 & 0 \\
 0 & 0 & 0 & 0 & 0 & 0 & 0 & 1  & 0 & 0 & 0 & 0 & 0 & 0 & 0 & 0 \\
 0 & 0 & 0 & 0 & 0 & 0 & 0 & 0 & 1  & 0 & 0 & 0 & 0 & 0 & 0 & 0 \\
 0 & 0 & 0 & 0 & 0 & 0 & 0 & 0 & 0 & 1  & 0 & 0 & 0 & 0 & 0 & 0 \\
 0 & 0 & 0 & 0 & 0 & 0 & 0 & 0 & 0 & 0 & 0 & 0 & 0 & 0 & 0 & 0 \\
 0 & 0 & 0 & 0 & 0 & 0 & 0 & 0 & 0 & 0 & 0 & 0  & 0 & 0 & 0 & 0 \\
 0 & 0 & 0 & 0 & 0 & 0 & 0 & 0 & 0 & 0 & 0 & 0 & 1  & 0 & 0 & 0 \\
 0 & 0 & 0 & 0 & 0 & 0 & 0 & 0 & 0 & 0 & 0 & 0 & 0 & 1  & 0 & 0 \\
 0 & 0 & 0 & 0 & 0 & 0 & 0 & 0 & 0 & 0 & 0 & 0 & 0 & 0 & 0  & 0 \\
 0 & 0 & 0 & 0 & 0 & 0 & 0 & 0 & 0 & 0 & 0 & 0 & 0 & 0 & 0 & 0  \\
\end{array}
\right).
\end{equation}
Notice that for the even power of \(\hat{V}_{AB}^{(int)}\), it has the form:
\begin{equation}
    [\hat{V}_{AB}^{(int)}]^n = \vert h \vert ^n \hat{D}.
\end{equation}
And so the odd power of n has the form:
\begin{equation}
    [\hat{V}_{AB}^{(int)}]^n= \vert h \vert ^{n-1} \hat{V}_{AB}^{(int)}.
\end{equation}
Therefore, the time evolution is given by:
\begin{equation}
\label{interaction}
    \hat{U}_{AB}^{(int)}(t) = \hat{I} - \hat{D} + \cos{(\Omega' t)} \hat{D} - i \frac{1}{\hbar \Omega'}\sin{(\Omega' t)} \hat{V}_{AB}^{(int)},
\end{equation}
where \(\Omega' = \vert h\vert/\hbar\). Expanding Eq. (\ref{interaction}) explicitly, we see that the time evolution takes the form given by Eq. (\ref{two atom unitary}) by setting a=0. 

We completed the first step of the proof, and now we want to show that we can construct a CZ gate for a two-atom unitary taking the form of Eq. (\ref{two atom unitary}). Following the same procedure of Ref. \cite{AQJ} with local unitaries redefined as:
\begin{equation}
    \hat{P}_1 = \left(
\begin{array}{cccc}
 1 & 0 & 0 & 0 \\
 0 & 1 & 0 & 0 \\
 0 & 0 & -\frac{i d}{| d| } & 0 \\
 0 & 0 & 0 & -\frac{i d}{| d| } \\
\end{array}
\right),
    \hat{P}_2 = \left(
\begin{array}{cccc}
 1 & 0 & 0 & 0 \\
 0 & 1 & 0 & 0 \\
 0 & 0 & \frac{i d^*}{| d| } & 0 \\
 0 & 0 & 0 & \frac{i d^*}{| d| } \\
\end{array}
\right)
\end{equation}
\begin{equation}
    \hat{P}_3 = \left(
\begin{array}{cccc}
 1 & 0 & 0 & 0 \\
 0 & 1 & 0 & 0 \\
 0 & 0 & \frac{e^{\frac{i \pi }{4}} \sqrt{c^* d^*}}{\sqrt{| c| } \sqrt{| d| }} & 0 \\
 0 & 0 & 0 & \frac{e^{\frac{i \pi }{4}} \sqrt{c^* d^*}}{\sqrt{| c| } \sqrt{| d| }} \\
\end{array}
\right),
\hat{P}_4 = \left(
\begin{array}{cccc}
 1 & 0 & 0 & 0 \\
 0 & 1 & 0 & 0 \\
 0 & 0 & \frac{e^{-\frac{i \pi }{4}} \sqrt{c d}}{\sqrt{| c| } \sqrt{| d| }} & 0 \\
 0 & 0 & 0 & \frac{e^{-\frac{i \pi }{4}} \sqrt{c d}}{\sqrt{| c| } \sqrt{| d| }} \\
\end{array}
\right),
\end{equation}
\begin{equation}
    \hat{P}_5 = \left(
\begin{array}{cccc}
 e^{-i \theta} & 0 & 0 & 0 \\
 0 & e^{-i \theta} & 0 & 0 \\
 0 & 0 & e^{i \theta} & 0 \\
 0 & 0 & 0 & e^{i \theta} \\
\end{array}
\right),
\hat{P}_6 = \left(
\begin{array}{cccc}
 1 & 0 & 0 & 0 \\
 0 & 1 & 0 & 0 \\
 0 & 0 & e^{2 i \theta} & 0 \\
 0 & 0 & 0 & e^{2 i \theta} \\
\end{array}
\right),
\end{equation}
where \(\theta = \vert c \vert + i \vert d\vert\).
\begin{equation}
\hat{S} = \left(
\begin{array}{cccc}
 1 & 0 & 0 & 0 \\
 0 & 1 & 0 & 0 \\
 0 & 0 & i & 0 \\
 0 & 0 & 0 & i \\
\end{array}
\right),
\hat{H} = \left(
\begin{array}{cccc}
 \frac{1}{\sqrt{2}} & 0 & \frac{1}{\sqrt{2}} & 0 \\
 0 & \frac{1}{\sqrt{2}} & 0 & \frac{1}{\sqrt{2}} \\
 \frac{1}{\sqrt{2}} & 0 & -\frac{1}{\sqrt{2}} & 0 \\
 0 & \frac{1}{\sqrt{2}} & 0 & -\frac{1}{\sqrt{2}} \\
\end{array}
\right),
\end{equation}
\begin{equation}
    \hat{Z} = \left(
\begin{array}{cccc}
 1 & 0 & 0 & 0 \\
 0 & 1 & 0 & 0 \\
 0 & 0 & -1 & 0 \\
 0 & 0 & 0 & -1 \\
\end{array}
\right).
\end{equation}
Perform the following gates in sequence:
\begin{equation}
    \hat{U}_2 = (\hat{P}_3 \otimes \hat{P}_3) (\hat{P}_1 \otimes \hat{P}_2) \hat{U}(t) (\hat{P}_3\otimes \hat{P}_4),
\end{equation}
\begin{equation}
    \hat{U}_3 = (\hat{S}^{\dag}\otimes \hat{S}^{\dag})( \hat{H} \otimes \hat{H}) \hat{U}_2 (\hat{H} \otimes \hat{H})(\hat{S} \otimes \hat{S}),
\end{equation}
\begin{equation}
    \hat{U}_4 = (\hat{I}_{4 x 4} \otimes \hat{Z})\hat{U}_3(\hat{I}_{4 x 4}\otimes \hat{Z}) \hat{U}_3,
\end{equation}
\begin{equation}
    \hat{U}_5 = (\hat{P}_6 \otimes \hat{P}_5) \hat{U}_4 = \left(
\begin{array}{cccccccccccccccc}
 1 & 0 & 0 & 0 & 0 & 0 & 0 & 0 & 0 & 0 & 0 & 0 & 0 & 0 & 0 & 0 \\
 0 & 1 & 0 & 0 & 0 & 0 & 0 & 0 & 0 & 0 & 0 & 0 & 0 & 0 & 0 & 0 \\
 0 & 0 & 1 & 0 & 0 & 0 & 0 & 0 & 0 & 0 & 0 & 0 & 0 & 0 & 0 & 0 \\
 0 & 0 & 0 & 1 & 0 & 0 & 0 & 0 & 0 & 0 & 0 & 0 & 0 & 0 & 0 & 0 \\
 0 & 0 & 0 & 0 & 1 & 0 & 0 & 0 & 0 & 0 & 0 & 0 & 0 & 0 & 0 & 0 \\
 0 & 0 & 0 & 0 & 0 & 1 & 0 & 0 & 0 & 0 & 0 & 0 & 0 & 0 & 0 & 0 \\
 0 & 0 & 0 & 0 & 0 & 0 & 1 & 0 & 0 & 0 & 0 & 0 & 0 & 0 & 0 & 0 \\
 0 & 0 & 0 & 0 & 0 & 0 & 0 & 1 & 0 & 0 & 0 & 0 & 0 & 0 & 0 & 0 \\
 0 & 0 & 0 & 0 & 0 & 0 & 0 & 0 & 1 & 0 & 0 & 0 & 0 & 0 & 0 & 0 \\
 0 & 0 & 0 & 0 & 0 & 0 & 0 & 0 & 0 & 1 & 0 & 0 & 0 & 0 & 0 & 0 \\
 0 & 0 & 0 & 0 & 0 & 0 & 0 & 0 & 0 & 0 & e^{4 i \theta} & 0 & 0 & 0 & 0 & 0 \\
 0 & 0 & 0 & 0 & 0 & 0 & 0 & 0 & 0 & 0 & 0 & e^{4 i \theta} & 0 & 0 & 0 & 0 \\
 0 & 0 & 0 & 0 & 0 & 0 & 0 & 0 & 0 & 0 & 0 & 0 & 1 & 0 & 0 & 0 \\
 0 & 0 & 0 & 0 & 0 & 0 & 0 & 0 & 0 & 0 & 0 & 0 & 0 & 1 & 0 & 0 \\
 0 & 0 & 0 & 0 & 0 & 0 & 0 & 0 & 0 & 0 & 0 & 0 & 0 & 0 & e^{4 i \theta} & 0 \\
 0 & 0 & 0 & 0 & 0 & 0 & 0 & 0 & 0 & 0 & 0 & 0 & 0 & 0 & 0 & e^{4 i \theta} \\
\end{array}
\right).
\end{equation}
By applying the conditions which \(\vert c \vert = \vert d\vert\), \(\hat{U}_5\) becomes a CZ gate. Specific to our example, \(\vert c \vert = \vert \cos{(\Omega' t)}\vert\), and \(\vert d \vert = \vert \frac{1}{\Omega'\hbar}\sin{(\Omega' t)}\vert \), the condition becomes that we require a time interval t satisfy:
\begin{equation}
    \Omega'^2\hbar^2 = \tan^2{(\Omega' t)}.
\end{equation}
Lastly, we want to comment that \(U_5\) can be viewed as four separate CZ gates coupling different pairs of degenerate states. For example, the first, third, ninth, and eleventh entries form a CZ gate between the degenerate ground state \(\vert\alpha_0\rangle, \vert\beta_0\rangle\), and the degenerate excited state \(\vert \alpha_2 \rangle, \vert \beta_2\rangle\). Similarly, the second, fourth, tenth, and twelfth entry forms the CZ gate between the degenerate ground state \(\vert\alpha_0\rangle, \vert\beta_1\rangle\) and the degenerate excited state \(\vert \alpha_2 \rangle, \vert \beta_3\rangle\), and so on. Therefore, under the assumption of identical atoms, the same interaction strength, and linearly polarized light-mediated coupling, it is postulated that performing the CZ gate is plausible.

\section{Higher Order Expansion of Degenerate Hadamard gate}
\begin{sidewaysfigure}
The first order correction is:
\begin{equation}
    \hat{U}_{Had}^{(1)} = 
    i\frac{\mu_BB_0}{\hbar\Omega}\left(
\begin{array}{cccc}
 -\frac{((\pi -4)(\hbar\omega/2)-30 \pi (\hbar\Omega/2)) \cos (\theta)}{12 \sqrt{2}(\hbar\omega/2)} & \frac{((\pi -4)(\hbar\omega/2)-30 \pi (\hbar\Omega/2)) \sin (\theta)}{12 \sqrt{2}(\hbar\omega/2)} & -\frac{\pi  ((\hbar\omega/2)-24(\hbar\Omega/2)) \cos
   (\theta)}{12 \sqrt{2}(\hbar\omega/2)} & \frac{\pi  ((\hbar\omega/2)-24(\hbar\Omega/2)) \sin (\theta)}{12 \sqrt{2}(\hbar\omega/2)} \\
 \frac{((\pi -4)(\hbar\omega/2)-30 \pi (\hbar\Omega/2)) \sin (\theta)}{12 \sqrt{2}(\hbar\omega/2)} & \frac{((\pi -4)(\hbar\omega/2)-30 \pi (\hbar\Omega/2)) \cos (\theta)}{12 \sqrt{2}(\hbar\omega/2)} & \frac{\pi  ((\hbar\omega/2)-24(\hbar\Omega/2)) \sin
   (\theta)}{12 \sqrt{2}(\hbar\omega/2)} & \frac{\pi  ((\hbar\omega/2)-24(\hbar\Omega/2)) \cos (\theta)}{12 \sqrt{2}(\hbar\omega/2)} \\
 \frac{\pi  (16(\hbar\Omega/2)+(\hbar\omega/2)) \cos (\theta)}{12 \sqrt{2}(\hbar\omega/2)} & -\frac{\pi  (16(\hbar\Omega/2)+(\hbar\omega/2)) \sin (\theta)}{12 \sqrt{2}(\hbar\omega/2)} & -\frac{(10 \pi (\hbar\Omega/2)+(\pi -4)(\hbar\omega/2)) \cos (\theta)}{12
   \sqrt{2}(\hbar\omega/2)} & \frac{(10 \pi (\hbar\Omega/2)+(\pi -4)(\hbar\omega/2)) \sin (\theta)}{12 \sqrt{2}(\hbar\omega/2)} \\
 -\frac{\pi  (16(\hbar\Omega/2)+(\hbar\omega/2)) \sin (\theta)}{12 \sqrt{2}(\hbar\omega/2)} & -\frac{\pi  (16(\hbar\Omega/2)+(\hbar\omega/2)) \cos (\theta)}{12 \sqrt{2}(\hbar\omega/2)} & \frac{(10 \pi (\hbar\Omega/2)+(\pi -4)(\hbar\omega/2)) \sin (\theta)}{12
   \sqrt{2}(\hbar\omega/2)} & \frac{(10 \pi (\hbar\Omega/2)+(\pi -4)(\hbar\omega/2)) \cos (\theta)}{12 \sqrt{2}(\hbar\omega/2)} \\
\end{array}
    \right),
\end{equation}

The second order correction is:
\begin{equation}
\scalemath{0.7}{
    \hat{U}_{Had}^{(2)} = \frac{(\mu_BB_0)^2}{(\hbar\Omega)^2}\left(
\begin{array}{cccc}
 -\frac{\pi  \left(900 \pi (\hbar\Omega/2)^2-60 (\pi -4)(\hbar\Omega/2)(\hbar\omega/2)+(\pi -4)(\hbar\omega/2)^2\right)}{288 \sqrt{2}(\hbar\omega/2)^2} & 0 & -\frac{576 \pi ^2(\hbar\Omega/2)^2-48 \pi ^2(\hbar\Omega/2)(\hbar\omega/2)+\left(16-4 \pi
   +\pi ^2\right)(\hbar\omega/2)^2}{288 \sqrt{2}(\hbar\omega/2)^2} & 0 \\
 0 & -\frac{\pi  \left(900 \pi (\hbar\Omega/2)^2-60 (\pi -4)(\hbar\Omega/2)(\hbar\omega/2)+(\pi -4)(\hbar\omega/2)^2\right)}{288 \sqrt{2}(\hbar\omega/2)^2} & 0 & -\frac{576 \pi ^2(\hbar\Omega/2)^2-48 \pi ^2(\hbar\Omega/2)(\hbar\omega/2)+\left(16-4
   \pi +\pi ^2\right)(\hbar\omega/2)^2}{288 \sqrt{2}(\hbar\omega/2)^2} \\
 -\frac{256 \pi ^2(\hbar\Omega/2)^2+32 \pi ^2(\hbar\Omega/2)(\hbar\omega/2)+\left(16-4 \pi +\pi ^2\right)(\hbar\omega/2)^2}{288 \sqrt{2}(\hbar\omega/2)^2} & 0 & \frac{\pi  \left(100 \pi (\hbar\Omega/2)^2+20 (\pi -4)(\hbar\Omega/2)
  (\hbar\omega/2)+(\pi -4)(\hbar\omega/2)^2\right)}{288 \sqrt{2}(\hbar\omega/2)^2} & 0 \\
 0 & -\frac{256 \pi ^2(\hbar\Omega/2)^2+32 \pi ^2(\hbar\Omega/2)(\hbar\omega/2)+\left(16-4 \pi +\pi ^2\right)(\hbar\omega/2)^2}{288 \sqrt{2}(\hbar\omega/2)^2} & 0 & \frac{\pi  \left(100 \pi (\hbar\Omega/2)^2+20 (\pi -4)(\hbar\Omega/2)
  (\hbar\omega/2)+(\pi -4)(\hbar\omega/2)^2\right)}{288 \sqrt{2}(\hbar\omega/2)^2} \\
\end{array}
\right)}.
\end{equation}
\end{sidewaysfigure}
\end{widetext}

\end{document}